\documentclass[aps,prd,reprint,preprintnumbers,showpacs,showkeys,superscriptaddress,nofootinbib,floatfix]{revtex4-1}
\usepackage{graphicx}
\usepackage{epsfig,amssymb,amsmath,color}
\usepackage{epstopdf}
\usepackage[colorlinks=true,citecolor=blue,linkcolor=blue,dvipdfmx]{hyperref}

\newcommand{\bea}{\begin{eqnarray}}   
\newcommand{\eea}{\end{eqnarray}}
\newcommand{\bear}{\begin{array}}  
\newcommand {\eear}{\end{array}}
\newcommand{\bef}{\begin{figure}}  
\newcommand {\eef}{\end{figure}}
\newcommand{\bec}{\begin{center}}  
\newcommand {\eec}{\end{center}}
\newcommand{\non}{\nonumber}  

\newcommand{\la}{\left\langle}
\newcommand{\ra}{\right\rangle}
\newcommand{\ds}{\displaystyle}

\def\EQ#1{Eq.~(\ref{#1})}

\def\GEV#1{10^{#1}{\rm\,GeV}}

\def\lrfp#1#2#3{ \left(\frac{#1}{#2} \right)^{#3}}

\newcommand{\sn}{{\rm sn}}
\newcommand{\cn}{{\rm cn}}

\begin{document}
\title{Elliptic Inflation: Interpolating from natural inflation to $R^2$-inflation}

\author{Tetsutaro Higaki}
\email{thigaki@post.kek.jp}
\affiliation{Theory Center, KEK, 1-1 Oho, Tsukuba, Ibaraki 305-0801, Japan}

\author{Fuminobu Takahashi}
\email{fumi@tuhep.phys.tohoku.ac.jp}
\affiliation{Department of Physics, Tohoku University, Sendai 980-8578, Japan}
\affiliation{Kavli Institute for the Physics and Mathematics of the
  Universe (WPI), Todai Institutes for Advanced Study, University of Tokyo,
  Kashiwa 277-8583, Japan}

\begin{abstract}
We propose an extension of natural inflation, where
the inflaton potential is a general periodic function.
{Specifically, we study elliptic inflation where the inflaton potential is given by 
Jacobi elliptic functions, Jacobi theta functions
or the Dedekind eta function, which appear in gauge and Yukawa couplings in the string theories 
compactified on toroidal backgrounds.
We show that in the first two cases }the predicted values of the spectral index and the tensor-to-scalar ratio 
interpolate from natural inflation to exponential inflation such as
$R^2$- and Higgs inflation and brane inflation, where the spectral index asymptotes to 
$n_s = 1-2/N \simeq 0.967$ for the e-folding number $N = 60$. 
{We also show that a model with the Dedekind eta function gives a sizable running of the spectral index due to 
modulations in the inflaton potential.} 
Such elliptic inflation can be thought of as a specific realization of 
multi-natural inflation, where the inflaton potential consists of multiple sinusoidal functions. 
We also discuss examples in string theory where Jacobi theta functions and the Dedekind eta function appear in the inflaton potential.
\end{abstract}
\preprint{KEK-TH-1788, TU-990, IPMU15-0003}
\maketitle

\section{Introduction}
\label{sec1}

The cosmic microwave background (CMB) is isotropic to roughly one part in $10^5$,
and its temperature anisotropies have the correlations beyond the horizon size
at the last scattering. This strongly implies that our Universe experienced an accelerated 
cosmic expansion, namely, inflation, at an early stage of the 
evolution~\cite{Guth:1980zm,Sato:1980yn,Starobinsky:1980te,Brout:1977ix,Kazanas:1980tx}.
In the slow-roll inflation paradigm~\cite{Linde:1981mu,Albrecht:1982wi},
the quantum fluctuations of the inflaton become scalar-mode metric perturbations (i.e. density perturbations),
while the quantum fluctuations of the graviton become tensor-mode metric perturbations~\cite{Starobinsky:1979ty}, 
inducing primordial $B$-mode polarization of the CMB. If the primordial $B$-mode polarization is detected, it will provide 
a definitive proof of inflation, and determine the inflation scale to be around the GUT scale. 

The inflaton field excursion during the last
$50$ or $60$ e-foldings must be greater than or comparable to the Planck scale, $M_P \simeq 2.4 \times \GEV{18}$,  in order to generate
large tensor mode  perturbations within the reach of the current and future CMB experiments.  
There have been proposed many such large-field inflation models.
Among them, there are two simple and therefore important
inflation models:  quadratic chaotic inflation~\cite{Linde:1983gd} and
natural inflation~\cite{Freese:1990rb}. 
The natural inflation consists of a single cosine function and it asymptotes to the quadratic chaotic inflation in the limit of a large decay constant $F$. 
The Planck constraint on the tensor-to-scalar ratio $r$ reads~\cite{Ade:2013uln}
\bea
r < 0.11~~(95\% {\rm~C.L.}),
\eea
and the quadratic chaotic inflation model is on the verge of being excluded, while some of the 
parameter space for the natural inflation $(F \lesssim 5 M_P)$ has been excluded.\footnote{
The BICEP2 result~\cite{Ade:2014xna} prefers a large $r$ which is consistent with the quadratic chaotic inflation. However,
the subsequent Planck polarization data~\cite{Adam:2014oea} showed that the dust contamination in the BICEP2 region is larger
than previously thought. 
}

There is one interesting class of inflation models where the predicted values of  $n_s$ and $r$ sits near the best-fit values of
the Planck observation: $R^2$ inflation~\cite{Starobinsky:1980te} or Higgs inflation~\cite{Bezrukov:2007ep} (see also
Refs.~\cite{Spokoiny:1984bd,Lucchin:1984yf,Goncharov:1983mw,Salopek:1988qh,Fakir:1990eg,Stewart:1994ts,Dvali:1998pa,Cicoli:2008gp} 
for related works). 
Of course this does not preclude any other models lying somewhere between the natural inflation and $R^2$ inflation. 
In fact, there are various extensions of the quadratic chaotic and natural inflation models, where a variety of values of $n_s$ and $r$ is realized.
 For instance,
if one considers a polynomial chaotic inflation with the potential $V(\phi) = \frac{1}{2} m^2 \phi^2 + \kappa \phi^3 + \lambda \phi^4 \cdots$,
the value of $r$ can be reduced to give a better fit to the Planck data~\cite{Nakayama:2013jka,Nakayama:2013txa,Kallosh:2014xwa,Nakayama:2014wpa}. 
Similarly, if there are multiple sinusoidal functions with different potential heights and decay constants, 
a variety of values of $n_s$ and $r$ can be 
realized~\cite{Czerny:2014wza,Czerny:2014xja,Czerny:2014qqa}, and such extension of the natural inflation is 
called {\it multi-natural inflation}. It was shown in Ref.~\cite{Czerny:2014wza} that, even with two sinusoidal functions, 
the lower bound on the decay constant disappears because a small-field inflation (axion hilltop inflation) 
is realized for a certain choice of the parameters. 
Note that, although multiple sinusoidal functions are generated in the  extra-natural inflation~\cite{ArkaniHamed:2003wu}, 
it does not significantly improve the bound on the decay constant as the higher order terms are suppressed.
It was recently pointed out that this problem can be circumvented  by adding appropriate matter fields in the bulk~\cite{Croon:2014dma}, 
and this provides another UV completion of the multi-natural inflation.

In this paper we propose another  extension of the natural inflation. The inflaton potential in natural inflation is given by
a single sinusoidal function, which is a specific type of periodic function. A periodic potential
is attractive from the model building point of view because it often appears when the flatness of the inflaton potential is protected  by an
approximate shift symmetry:
the inflaton potential which arises from the breaking of a continuos shift symmetry  respects the residual discrete one,
leading to such  periodic potential.  We therefore focus on a class of inflation models with periodic potential. 
{Specifically, we consider Jacobi elliptic functions, Jacobi theta functions, and the Dedekind eta function }as the inflaton potential, and 
we call such inflation models based on elliptic functions and other related functions as {\it elliptic inflation}.

\vspace{2mm}

 As an example of the elliptic inflation, we will first consider
a Jacobi  elliptic function as the inflaton potential: 
\bea
V(\phi) = 2 \Lambda^4 \left(1-{\rm cn}^2\left( \frac{\phi}{2 F},k\right)\right),
\label{Jacobi}
\eea
where ${\rm cn}(x,k)$ is the Jacobi  elliptic cosine
function with elliptic modulus $0 \leq k \leq 1$, and  $\Lambda$ and $F$ are model parameters.
We call the inflation model with the above potential {\it cnoidal inflation}.
Interestingly,  the cnoidal inflation is reduced to the natural inflation in the limit of $k \to 0$, while it is reduced
to the exponential inflation including the brane inflation, 
$R^2$ inflation, and the Higgs inflation, in the limit of $k \to 1$.

The Jacobi elliptic functions are closely related to the Jacobi theta functions,
which often appear in theories with periodic boundary conditions.
{Periodicity or modular invariance play important roles in field theories in association with symmetries and dualities~\cite{Seiberg:1994rs,Seiberg:1994aj}.}
In  string theories, periodicity appears through compactifications of extra dimensions when a compact manifold has periodic properties.
For instance, a holomorphic gauge coupling at one loop level 
is given by the Jacobi theta function $\vartheta_1 (v,q)$ in toroidal compactification~\cite{Berg:2004ek}. 
%
%
Here, $v$ is a open string modulus (diagonal modes of the adjoint representation 
on D-branes), 
and $q = e^{i\pi \tau}$ is given by a complex structure of a torus $\tau$ in type IIB models
(see also Ref.~\cite{Baumann:2006th} for their interpretation and further generalization).
In our later example, the inflaton is identified with the real part of $v$.
Another example is a Yukawa coupling among matter-like fields, which can be described 
by the Jacobi theta functions $\vartheta_3 (v,q)$
in magnetized (or intersecting) D-brane models on tori \cite{Cremades:2003qj,Cremades:2004wa}.\footnote{
In both IIB and IIA models on tori, 
it is possible to realize magnetized and intersecting D-brane system with any branes
if one does not require the preservation of supersymmetry \cite{Marino:1999af}.
} 
%
%
The theta functions appear in the gauge and Yukawa couplings because of the periodicity of the tori.
Similar Yukawa couplings are obtained in
Heterotic orbifolds through world-sheet instantons \cite{Hamidi:1986vh}.\footnote{
In Ref.\cite{Dundee:2010sb}, they considered a case in which Yukawa couplings are given by 
the Dedekind eta function of K\"ahler moduli with the target space modular invariance.
}
Therefore, a scalar potential which depends on the theta functions appears naturally, and
we shall show that it takes the following form,
\bea
V(\phi) \sim  \bigg[ {\vartheta}_1 \bigg(\frac{\phi}{2\pi F}, q \bigg) \bigg]^{a} 
\bigg[ {\vartheta}_3 \bigg(\frac{\phi}{2\pi F}, q \bigg) \bigg]^{b},
\label{thetaab}
\eea
where $a$ and $b$ denote rational numbers. We call the inflation model with the
above potential form {\it theta inflation}.
We find that the theta inflation exhibits similar behavior to the cnoidal inflation. 


{
It is possible to consider inflation models based on other special functions.
In particular, inflation models with the Dedekind eta function was studied in Ref.~\cite{Abe:2014xja}.
(See also Ref.~\cite{Schimmrigk:2014ica} for inflation models with modular functions.)
Interestingly,  the inflaton potential can be approximated by natural inflation with modulations, 
leading to a sizable running of the spectral index.
In fact, it was shown in Ref.~\cite{Kobayashi:2010pz} that a sizable running of the spectral
index can be generated if there are small modulations on the inflaton potential.
(See also Refs.~\cite{Feng:2003mk,Takahashi:2013tj,Czerny:2014wua,Abazajian:2014tqa,Wan:2014fra,Minor:2014xla,delaFuente:2014aca} for related work.)
Here we will study the inflaton potential with the eta function in a greater detail.
}


\vspace{2mm}

The rest of this paper is organized as follows. In Sec.~\ref{sec2} we study phenomenological 
aspects of the elliptic inflation, where we calculate the predicted values of $n_s$ and $r$
for the above three elliptic inflation models. In Sec.~\ref{sec3} we discuss possible origins of the Jacobi
theta functions and the Dedekind eta function in the string theories and implications for the large-field inflation. 
The last section
is devoted to discussion and conclusions. 

\section{Elliptic inflation}
\label{sec2}
In this section we first study two examples of the elliptic inflation. One is based on the Jacobi
elliptic inflation (cnoidal inflation), and the other is based on the Jacobi theta function (theta
inflation). As we shall see below, they provide a one-parameter extension of the
natural inflation,  interpolating from natural inflation (approximately) to the $R^2$-inflation. 
{Finally we study the inflation based on the Dedekind eta function, and show that a
sizable running of the spectral index can be generated because of the small modulations 
on the inflaton potential.}

\subsection{Cnoidal inflation}
\begin{figure}[t!]
\begin{center}
\includegraphics[width=7cm]{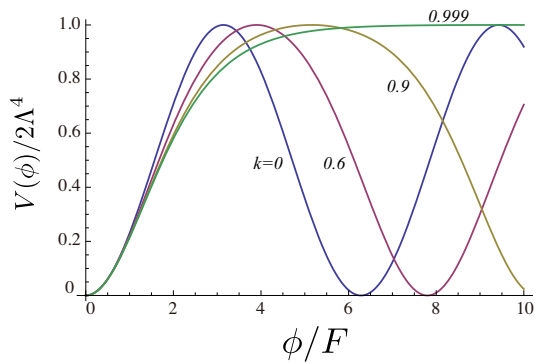}
\caption{
The inflaton potential $V(\phi)$ with the Jacobi elliptic function given by \EQ{Jacobi} for $k=0, 0.6, 0.9$ and $0.999$.
}
\label{cinoidalV}
\end{center}
\end{figure}

Let us  consider the cnoidal inflation model with the potential \EQ{Jacobi}. The Jacobi  elliptic functions $\sn(x,k)$ and $\cn(x,k)$ can
be thought of as an extension of trigonometric functions, $\sin(x)$ and $\cos(x)$, respectively. In the limit of $k \to 0$, they 
become
\bea
\sn(x,k) & \to & \sin x,\\
\cn(x,k) & \to & \cos x,
\eea
while, in the limit of $k \to 1^-$, 
\bea
\sn(x,k) & \to & \tanh x,\\
\cn(x,k) & \to & {\rm sech} \,x.
\eea
In these limits, the inflaton potential $V(\phi)$ becomes
\bea
V(\phi) & \to &
\left\{
\bear{l}
\ds{\Lambda^4 \left(1 - \cos \frac{ \phi}{F} \right)}  {\rm~~~~for~~}k\to 0,\\
\ds{
2\Lambda^4 \tanh^2\frac{\phi}{2 F} \simeq 
2\Lambda^4 \left(1-4 e^{-\phi/F}\right)} \\
~~~~~~~~~~~~~~~~~~~~~ {\rm~~~for~~}k\to 1,\\
\eear
\right.
\eea
where we have assumed $\phi \gg F$ in the second equality.  Therefore, the elliptic inflaton potential \EQ{Jacobi} interpolates
from natural inflation to the exponential inflation. See Fig.~\ref{cinoidalV} for the inflaton potential $V(\phi)$ for several values of $k$.
One can see from the figure that the potential becomes flatter around the potential maximum as $k$ approaches unity.

To see how the predicted values of the spectral index $n_s$ and the tensor-to-scalar ratio $r$ change as a function of the model parameters, 
we numerically solve the inflaton equation of motion,
\bea
\ddot{\phi} + 3H \dot{\phi} + V'(\phi) =0,
\eea
with $V(\phi)$ given by \EQ{Jacobi}. The spectral index and the tensor-to-scalar ratio can be evaluated by the
standard formulae,
\bea
n_s & \simeq & 1 + 2 \eta - 6 \varepsilon,\\
r &\simeq & 16 \varepsilon,
\eea
where $\varepsilon$ and $\eta$ are the slow-roll parameters,
\bea
\varepsilon &=& \frac{1}{2} \lrfp{V'(\phi)}{V(\phi)}{2},\\
\eta &=& \frac{V''(\phi)}{V(\phi)}.
\eea
Here and in what follows we have adopted the Planck units, $M_P = 1$, and the prime denotes the derivative with respect to the
inflaton field $\phi$. In the actual numerical calculation, we have used a refined version of the above formulae including
up to the second-order slow-roll parameters.

\begin{figure}[t!]
\begin{center}
\includegraphics[width=7cm]{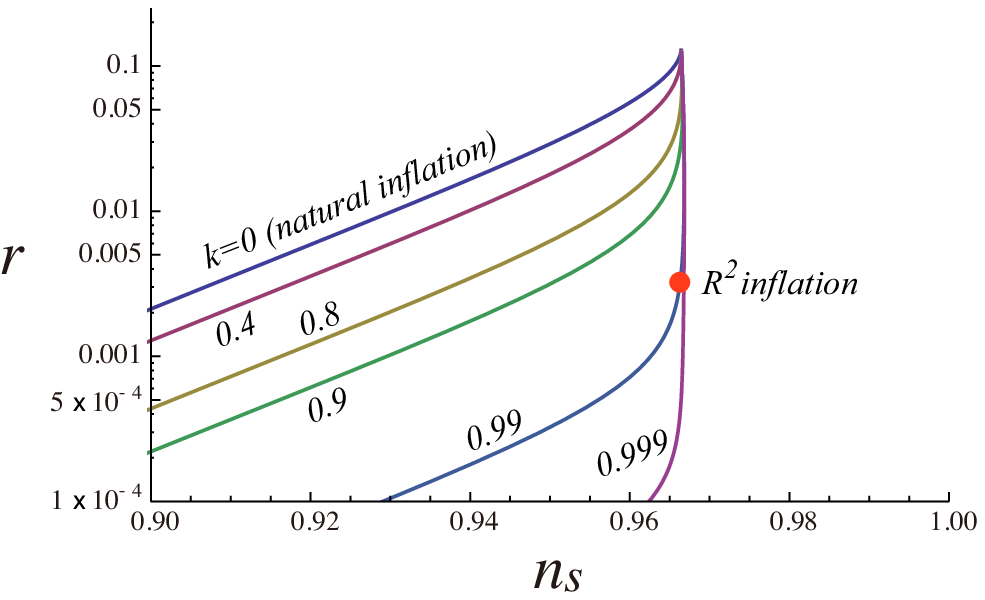}
\includegraphics[width=7cm]{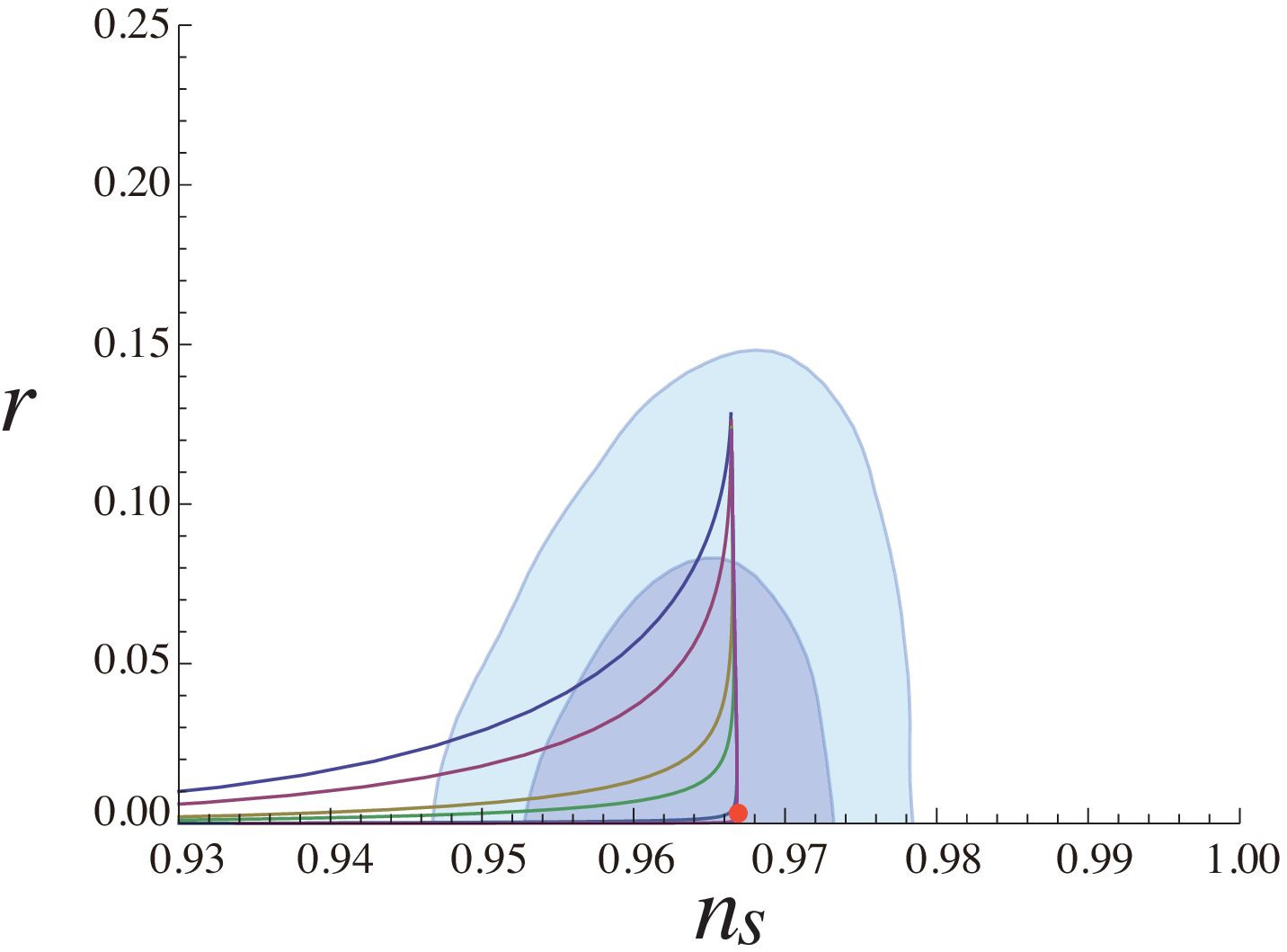}
\caption{The spectral index and the tensor-to-scalar ratio in the elliptic inflation \EQ{Jacobi} for $k = 0, 0.4, 0.8, 0.99$ and $0.999$ 
from top to bottom with the e-folding number $N=60$.
The prediction of $R^2$-inflation is shown by a red disc. 
We also overlay the Planck result ({\it Planck}+WP+BAO)~\cite{Ade:2013uln} in the bottom panel for comparison. 
The elliptic inflation interpolates from natural inflation to $R^2$-inflation as $k$ varies from $0$ to $1$.
}
\label{cinoidal}
\end{center}
\end{figure}

\begin{figure}[th]
\begin{center}
\includegraphics[width=7cm]{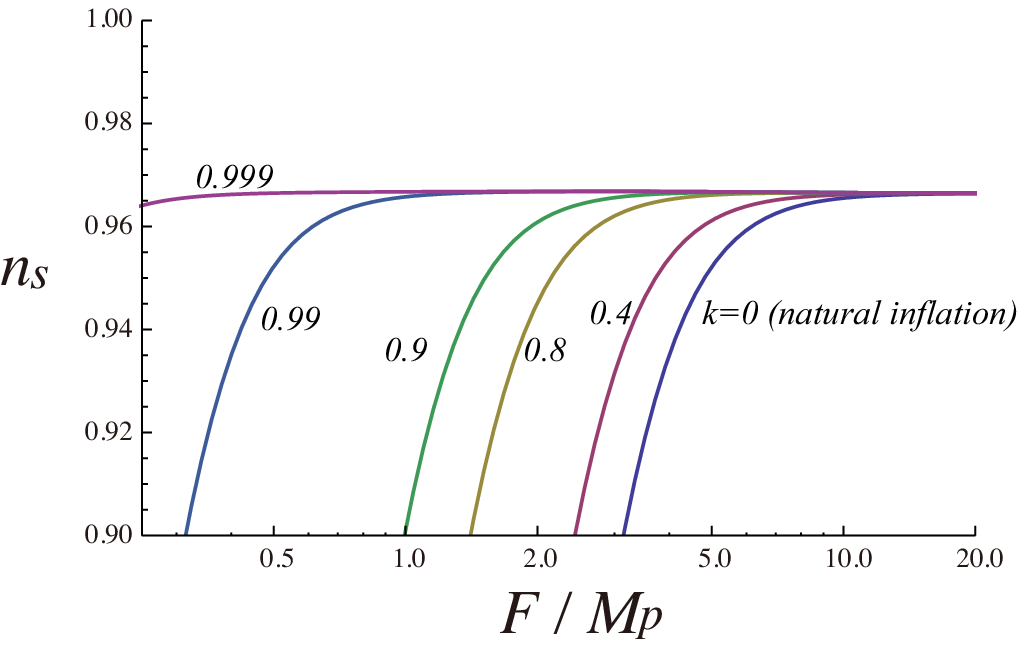}
\includegraphics[width=7cm]{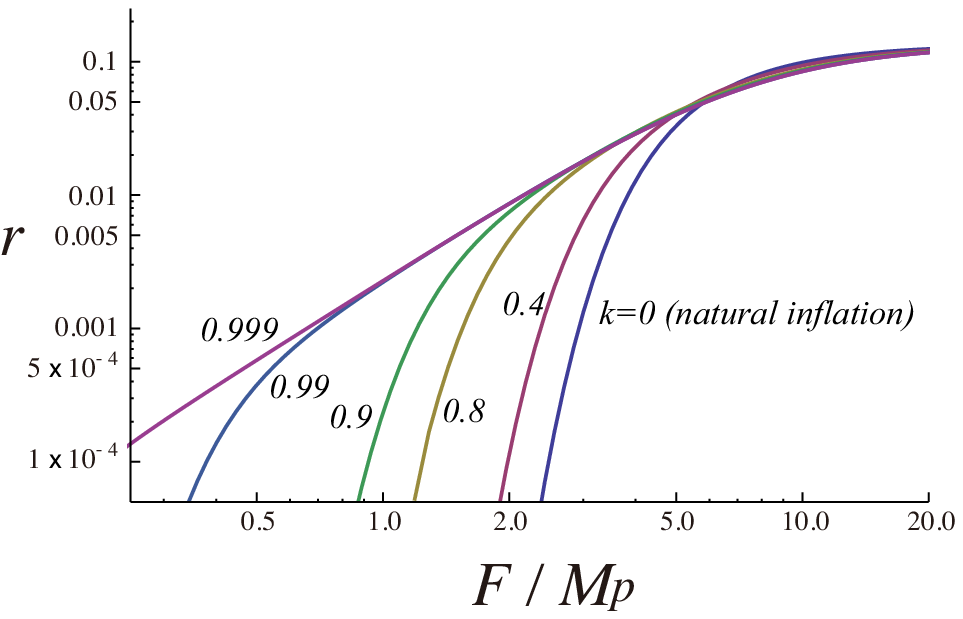}
\caption{
The dependence of $n_s$ and $r$ on the decay constant $F$ in the cnoidal inflation \EQ{Jacobi}.
}
\label{cinoidal2}
\end{center}
\end{figure}

In Fig.~\ref{cinoidal} we show the calculated values of $n_s$ and $r$ for several values of $k$ by varying $F$. One can see
that the $(n_s, r)$ predicted in the elliptic inflation indeed interpolates from natural inflation to exponential inflation.
The latter includes the $R^2$- or Higgs inflation, in which the potential is given by 
$V(\phi) \sim (1-2 e^{-\sqrt{\frac{2}{3}}\phi})$. Here we have fixed the e-folding number to be $N=60$. 
We also show in Fig.~\ref{cinoidal2} how the $n_s$ and $r$ evolve as the decay constant $F$ changes.
For $k$ sufficiently close to unity ($k \gtrsim 0.99$), the spectral index $n_s$ stays around 
$1-2/N = 0.967$ for a wide range of the decay constant
including even sub-Planckian values.

\subsection{Theta inflation}
Next we consider another example of the elliptic inflation,  the theta inflation with the potential (\ref{thetaab}).
It is known that the Jacobi  elliptic functions are closely related to the Jacobi theta functions, $\vartheta_{0,1,2,3}(v, \tau)$.
For instance, there are following relations:
\bea
\sn(u,k) &=& \frac{\vartheta_3(0,q) \,\vartheta_1(v,q)}{\vartheta_2(0,q)\, \vartheta_0(v,q )},\\
\cn(u,k) &=& \frac{\vartheta_0(0,q)\, \vartheta_2(v,q)}{\vartheta_2(0,q) \, \vartheta_0(v,q )},
\eea
where
\bea
k &=& \bigg(\frac{\vartheta_2(0,q)}{\vartheta_3(0,q)} \bigg)^2,~~~ u =  \pi v (\theta_3(0,q))^2 ,
\eea
or equivalently,
\bea
v &=& \frac{u}{2K(k)},~~q =  e^{i \pi \tau},~~
\tau = \frac{i K(\sqrt{1-k^2})}{K(k)},
\eea
and $K(k)$ is the elliptic integral of the first kind. The Jacobi theta functions are given by
\bea
\label{theta0}
\vartheta_0(v,q) &=&
\sum_{n=-\infty}^{\infty}e^{\pi i \tau n^2 + 2\pi i n \big(v +\frac{1}{2} \big)} \non\\
&=& 1+2\sum_{n=1}^\infty(-1)^n q^{ n^2} \cos 2n\pi v,
\label{theta1}
\eea
\bea
\vartheta_1(v,q) &=& 
\sum_{n=-\infty}^{\infty}e^{\pi i \tau \big(n + \frac{1}{2}\big)^2 + 2\pi i \big(n +\frac{1}{2} \big) \big(v +\frac{1}{2} \big)} \non\\
&=&2\sum_{n=0}^\infty(-1)^n q^{ \left(n+\frac{1}{2}\right)^2} \sin \left(2n+1\right) \pi v,
\label{theta2}
\eea
\bea
\vartheta_2(v,q) &=& 
\sum_{n=-\infty}^{\infty}e^{\pi i \tau \big(n + \frac{1}{2}\big)^2 + 2\pi i \big(n +\frac{1}{2} \big) v} \non\\
&=&2\sum_{n=0}^\infty q^{ \left(n+\frac{1}{2}\right)^2} \cos \left(2n+1\right) \pi v,
\eea
\bea
\vartheta_3(v,q) &=& 
\sum_{n=-\infty}^{\infty}e^{\pi i \tau n^2 + 2\pi i n v} \non\\
&=& 1+2\sum_{n=1}^\infty q^{ n^2} \cos 2n \pi v.
\label{theta3}
\eea
Here, $|q| <1$, i.e., Im$(\tau) >0$.
The Fourier series converge very fast for $|q| \ll 1$ as the higher order terms are exponentially suppressed. 

The Jacobi theta functions  appear in gauge and Yukawa couplings 
in string theories with toroidal compactifications. As we shall see in the next section,
 the Jacobi theta functions  appear naturally in the scalar potential.
 
\begin{figure}[t!]
\begin{center}
\includegraphics[width=7cm]{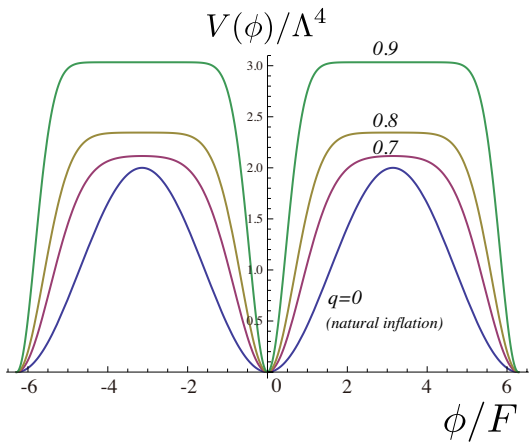}
\caption{
The inflaton potential $V(\phi)$  with the Jacobi theta function given by \EQ{theta} for $q=0, 0.7, 0.8$ and $0.9$ from
bottom to top.
}
\label{ThetaV}
\end{center}
\end{figure}

To be specific, let us consider the following  inflaton potential,
\bea
V(\phi) &=&
\frac{\Lambda^4}{2q} \left( 
\vartheta_3(0,q) - \vartheta_3\left(\frac{\phi}{2 \pi F},q \right)
\right)
\label{theta}
\eea
which corresponds to Eq.~(\ref{thetaab}) with $a = 0$ and $b=1$. 
Note that a similar inflaton dynamics is obtained for the potential with the other theta functions
because of the identities between the theta functions: 
$\vartheta_3(v+1/2,q)=\vartheta_0(v,q)$, $\vartheta_3(v+\tau/2,q) =
e^{-i\pi(\tau/4+v)}\vartheta_2(v,q)$, $\vartheta_2(v+1/2,q) =
-\vartheta_1(v,q)$ and so on.
In the limit of $q \to 0$, the above potential is reduced to the natural inflation since the higher order
terms in the Fourier series are suppressed. 
\bea
V(\phi) \to \Lambda^4 \left(1-\cos\frac{\phi}{F} \right)~~~~{\rm for~~}q \to 0.
\eea
On the other hand, for $|q| \to 1$, many sinusoidal functions add up so as to make the inflaton potential 
extremely flat around the potential maximum. This is essentially same as the axion hilltop inflation where the inflaton potential becomes
flat due to cancellation among  multiple sinusoidal functions~\cite{Czerny:2014wza,Czerny:2014xja,Czerny:2014qqa}.
In this sense, the theta inflation  can be thought of as a specific realization of the multi-natural inflation. (Also note that the
Fourier series converge fast unless $|q|$ is extremely close to unity.)
We show the inflaton potential for 
$q=0, 0.7, 0.8$ and $0.9$
in Fig.~\ref{ThetaV}.
As one can see from the figure, the inflaton potential around the local maximum becomes 
flatter as $q$ approaches unity.  

\begin{figure}[t!]
\begin{center}
\includegraphics[width=7cm]{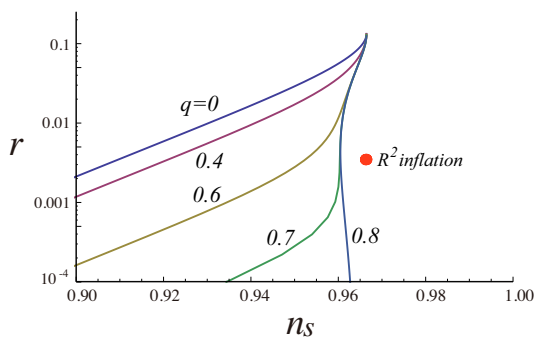}
\caption{
$(n_s, r)$ for the theta inflation with the potential (\ref{theta}).
}
\label{fig:theta}
\end{center}
\end{figure}

\begin{figure}[t!]
\begin{center}
\includegraphics[width=7cm]{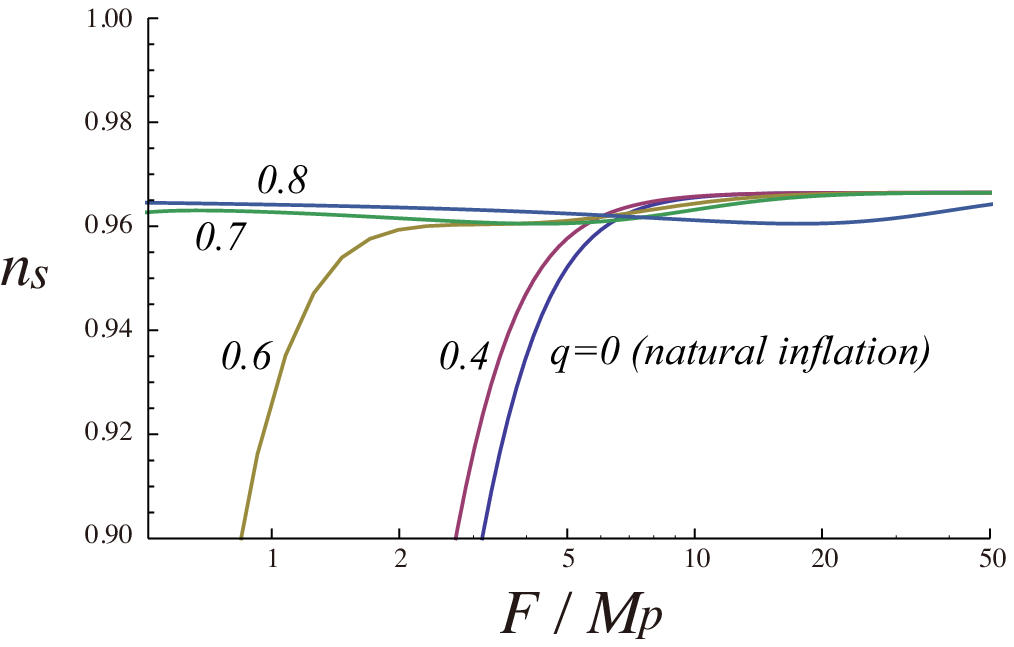}
\includegraphics[width=7cm]{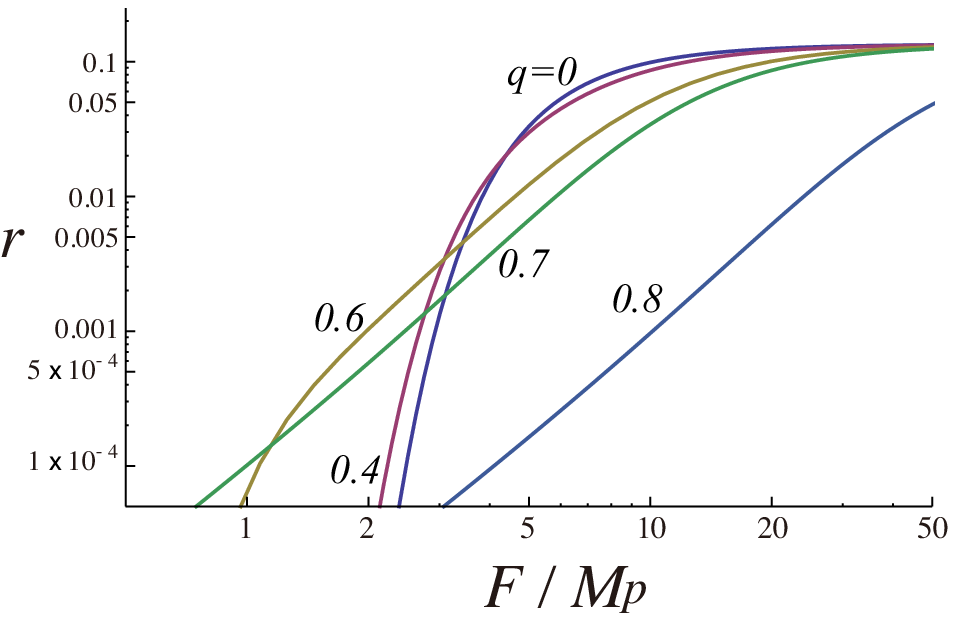}
\caption{
Same as Fig.~\ref{cinoidal2} but for the theta inflation (\ref{theta}).
}
\label{fig:theta2}
\end{center}
\end{figure}

We similarly solve the inflaton dynamics with inflaton potential (\ref{theta}), and estimated
$n_s$ and $r$. The results are shown in Figs.~\ref{fig:theta} and \ref{fig:theta2}. The spectral index $n_s$
stays around $0.967$ for sub-Planckian values of $F$, if $q \gtrsim 0.7$.
It is interesting to note that 
the predicted values of $n_s$ and $r$ become closer to the $R^2$-inflation as $q$ becomes larger,
but they do not exactly matches with  the $R^2$-inflation: the spectral index tends to be slightly
smaller  for the same value of $r$.

\subsection{Inflation with Dedekind eta function}

Now we study an inflation model with the Dedekind eta function, 
\bea
\eta (\tau) = q^{\frac{1}{12}}\prod_{n=1}^{\infty}(1-q^{2n}) ,
\eea
where $q = e^{i \pi \tau}$ and ${\rm Im}(\tau) > 0$.
It is known that the eta function is given by the special values of the theta functions,
for instance,
$\eta^3 (\tau) = 
\frac{1}{2} \vartheta_2(0,\tau) \vartheta_3 (0,\tau) \vartheta_0 (0,\tau) .$

\begin{figure}[t!]
\begin{center}
\includegraphics[width=7cm]{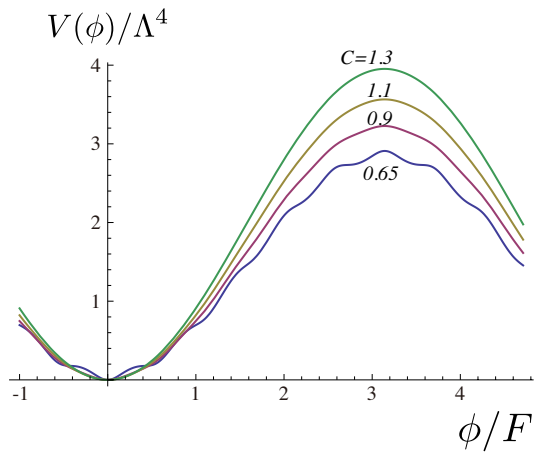}
\caption{
The inflaton potential with the Dedekind eta function (\ref{Veta}) for
$C = 1.3, 1.1, 0.9$ and 0.65 from top to bottom.
}
\label{fig:dedekindV}
\end{center}
\end{figure}

To be specific, we will study the inflaton potential:
\bea
\label{Veta}
V(\phi) &=&- \Lambda^4 \left[
\eta^{-2} \bigg(\frac{6}{\pi F}\phi + i C \bigg) + c.c. \right]
+{\rm const.}, \\
&\propto& \cos \frac{\phi}{F} + e^{-2\pi C} \cos \left(11 \frac{ \phi}{F}\right) + \cdots,\non
\eea
where $C$ is a real parameter, and the constant term is added so as to make the inflaton potential vanish at the origin, 
 $V(0) = 0$. In the second line, we have expanded the eta function in terms of  $|q| = e^{- \pi C} \ll 1$.
The inflaton potential (\ref{Veta}) is shown
in Fig.~\ref{fig:dedekindV} for several values of $C$. One can see from the figure that the inflaton 
potential has a maximum at $\phi = \pi F$ and it receives modulations for $C \lesssim 1$.
Here, $(\phi, C) = (0,1)$ implies the self-dual point under the modular transformation
of $\tau \to -1/\tau$, i.e., $C \to 1/C$.
For larger values of $C$, the above potential asymptotes to the 
natural inflation with the decay constant $F$, as one can see from the expansion of the inflaton potential.
Note that the normalization of the potential
height is chosen for visualization purpose. This does not affect the predicted
values of $r$, $n_s$ and its running in the following analysis.

\begin{figure}[t!]
\begin{center}
\includegraphics[width=7cm]{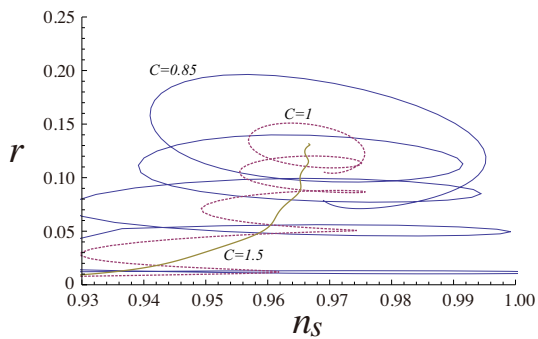}
\caption{
$(n_s, r)$ for the inflation  potential with the Dedekind eta function (\ref{Veta}).
}
\label{fig:dedekindnsr}
\end{center}
\end{figure}

\begin{figure}[t!]
\begin{center}
\includegraphics[width=7cm]{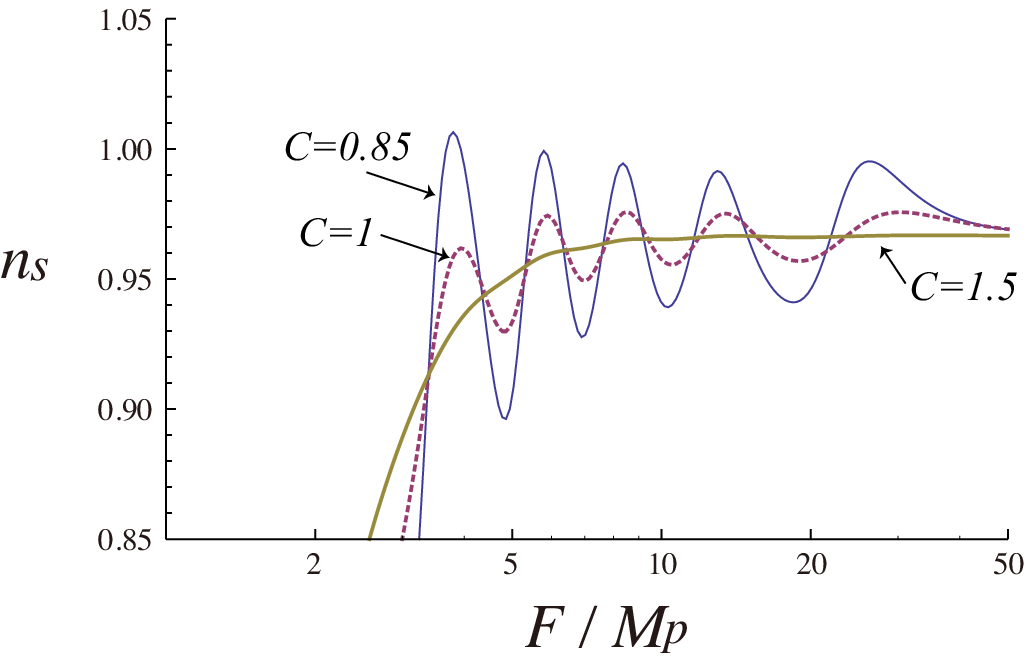}
\includegraphics[width=7cm]{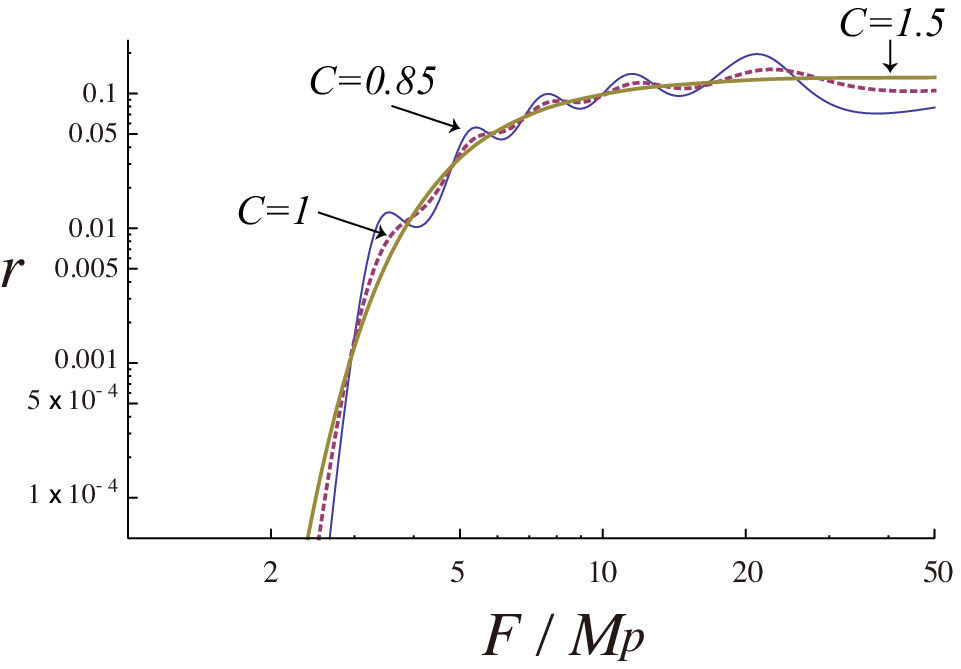}
\caption{
Same as Fig.~\ref{cinoidal2} but for the inflaton potential (\ref{Veta}).
}
\label{fig:dedekindnsrF}
\end{center}
\end{figure}

The predicted values of $n_s$ and $r$ are shown in Fig.~\ref{fig:dedekindnsr}
for $C = 1.5, 1$ and $0.85$, 
where we have varied the value of the decay constant $F$, and set
the e-folding number $N$  to be $60$. For $C = 1$ and $C = 0.85$, the predicted values of $(n_s, r)$ rotate
counterclockwise as $F$ decreases.
In contrast to the previous two cases, this model does not asymptote to
the exponential inflation. The dependence of $n_s$ and $r$ 
on the decay constant $F$ is shown in Fig.~\ref{fig:dedekindnsrF}, and we find
that the super-Planckian decay constant is required to be consistent with
the Planck data. We will return to the issue of super-Planckian decay constant
at the end of this section.

\begin{figure}[t!]
\begin{center}
\includegraphics[width=7cm]{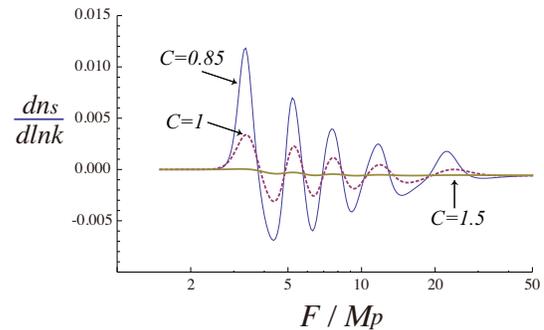}
\caption{
The running of the spectral index $d n_s/d\ln k$ as a function of the decay constant $F$ 
in the inflation model  (\ref{Veta}).
}
\label{fig:dedekindnsalpha}
\end{center}
\end{figure}

Finally let us evaluate the running of the spectral index, $dn_s/d \ln k$.
Recently the inflation model with the Dedekind eta
function was shown to generate a sizable running of the spectral index~\cite{Abe:2014xja}.
The running can be expressed in terms of the slow-roll parameters as
\bea
\frac{dn_s}{dlnk} \simeq -24\varepsilon^2 +16\varepsilon \eta - 2\xi ,
\label{running}
\eea
where
\bea
 \xi \equiv \frac{V'(\phi)V'''(\phi)}{V(\phi)^2}.
\eea
 The current constraint
on the running by {\it Planck} and WMAP polarization reads~\cite{Ade:2013zuv}
\bea
\frac{dn_s}{d\ln k} = -0.013 \pm 0.009 ~~(68\%).
\eea
For $\varepsilon$ and $\eta$ of $O(0.01)$, the first two terms in Eq.~(\ref{running})
are of $O(10^{-3})$, one order of magnitude smaller than the center value of the current bound.
In order to generate a sizable running of order $O(0.01)$, therefore,  $\xi$ must be 
large. For a simple class of inflaton potentials expressed by Taylor series with the finite truncation, however, 
a large negative running that is more or less constant over the observed cosmological
scales would quickly terminate inflation within 
the e-folding number $N \lesssim 30$~\cite{Easther:2006tv}. There is a loophole in this analysis, and
it was pointed out in Ref.~\cite{Kobayashi:2010pz}
that a sizable and almost constant running over the CMB scales $\Delta N_{\rm CMB} \sim 8$ can be generated
in an inflaton potential with small modulations, because the modulations can give a dominant contribution
to $V'''$, while $V$, $V'$ and $V''$ are not significantly modified. 
The running of the spectral index as a function of the decay constant is shown in Fig.~\ref{fig:dedekindnsalpha}. 
For slightly smaller values of $C$, $|dn_s/d\ln k|$ can be larger than $0.01$, and it is more or less constant
over the CMB scales. 

\section{Possible UV completion}
\label{sec3}

We have introduced the Jacobi  elliptic functions and theta functions in the scalar potential to extend 
the natural inflation, motivated by the recent Planck data.
We have shown that they predict the spectral index and the tensor-to-scalar ratio interpolating from natural inflation
to exponential inflation including the $R^2$- or Higgs inflation. Such functions indeed appear in the string
theories with extra dimension containing tori. As noted already, gauge and Yukawa interactions 
can be written in terms of the Jacobi theta functions in toroidal compactifications 
in intersecting or magnetized D-brane system~\cite{Berg:2004ek,Cremades:2003qj,Cremades:2004wa}.
Also on the magnetized orbifolds, e.g., $T^6/ ({\mathbb Z}_2\times {\mathbb Z}_2)$ and
$T^2 \times T^4/ {\mathbb Z}_2$,
Yukawa couplings are similarly expressed with the theta functions \cite{Abe:2014vza}.
This type of function comes from the wave functions of the fields on the torus, 
which respects the periodicity, in magnetized brane models
(or summation of world-sheet instantons in intersecting brane models).\footnote{
The wave functions are invariant under the periodic shift on the torus 
up to gauge transformation, which is represented by double pseudo periodicity of the Jacobi theta functions. 
}
So, the scalar potential with the Jacobi theta function will appear naturally in the low energy scales.

In passing, we note that the theta functions  can also appear in scattering amplitudes
at one-loop level of string world-sheet together with the Dedekind eta function (involving 
moduli integrals) \cite{Polchinski:1998rq}. This is similar in the field theories \cite{Antoniadis:2001cv}.
Also the eta function can appear
in one-loop threshold corrections to gauge couplings 
after integrating out heavy Kaluza-Klein modes in the extra dimension \cite{Blumenhagen:2006ci}.
Note that the eta function is given by the special value of the theta functions, as we have seen before.

In the following, we  discuss several possibilities to generate
the scalar potential with the theta functions.
The inflaton candidate is an open string modulus on the D-branes, and hence it
originates from the Wilson lines, positions of D-branes or their linear combination.
For instance, if a four cycle ${\cal S}$ in a Calabi-Yau, wrapped by D7-branes,
has the Hodge numbers with $h^{1,0}_-({\cal S}) \neq 0$ or $h^{2,0}_-({\cal S}) \neq 0$, 
there appear $h^{1,0}_-({\cal S})$ Wilson line moduli or 
$h^{2,0}_-({\cal S}) $ position moduli, respectively \cite{Jockers:2004yj}. 
To discuss whether such models are really viable for inflation, however,
it is necessary to construct an explicit model of the moduli stabilization, 
supersymmetry (SUSY) breaking, as well as the embedding of the standard model sector etc. 
The open string inflation and the related issues have been studied extensively in the literature.
The purpose of this section is to present several possible scenarios
 where the Jacobi theta functions  appear in the scalar potential, motivated by the
recent Planck results on $n_s$ and $r$.

\subsection{Field theoretic cases}
As a toy model, we first consider a case where the Yukawa coupling is given by the theta function
$\vartheta_3(v,q)$, where $q=e^{i \pi \tau}$, and $\tau$ is a torus complex structure modulus $\tau$
in magnetized brane model~\footnote{
A Yukawa coupling consists of a product of theta functions 
if a compact extra dimension is given by a product of tori. 
We will focus on the contribution from a single torus and
neglect contributions from the other tori, assuming that other moduli become heavy owing to flux compactification. One may consider $T^4/{\mathbb Z}_{2} \times T^2$, in which
Wilson line moduli could be required to explain the smallness of the observable Higgs mass,
although the moduli are not necessary to explain Yukawa textures \cite{Abe:2014vza}.
On the other hand, 
$\vartheta_3$ is often used for realizing a large (top) Yukawa coupling because it contains unity and 
$\vartheta_2$ is used for a smaller top Yukawas; see Eq.~(\ref{theta3}). }
(or the volume modulus and B-field in the intersecting brane system), and $v$ is identified with the inflaton $\phi/2\pi F$. 
In a more realistic scenario, one has to worry about the stabilization of the imaginary component of $v$,
which we shall discuss in the next subsection.\footnote{Note also that theta inflation with $|q| \sim 1$ would require
a small complex structure, i.e., ${\rm Im}(\tau) \ll 1$.
Then we may have a difficulty to control the coupling expansion,
because then a complex structure on a torus can be translated into the volume under T-duality.
{Such a problem may be avoided,
if $q$ in the theta function depends on a linear combination of moduli 
or a large volume extra dimension is realized with stabilized moduli, compensating small values 
of complex structure moduli.}
}

Now the question is how to generate a scalar potential which depends on the Yukawa coupling. 
One possibility is to use a strong dynamics:
\bea
\nonumber
{\cal L} &=&  {\vartheta}_3 (v,q ) \varphi \psi^c \psi  \\
 ~ \to ~V&=&  {\vartheta}_3 (v,q)  \langle \varphi \rangle
\langle \psi^c \psi \rangle \equiv \Lambda^4 {\vartheta}_3 \bigg(\frac{\phi}{2\pi F} ,q\bigg),
\eea
where $\varphi$ and $\psi + \psi^c$ are scalar and vector-like fermion fields 
in the strongly coupled hidden sector respectively.
Here, we have assumed that the scalar develops a non-zero 
vacuum expectation value (VEV) and the fermion pair condensates through the strong dynamics.

In a supersymmetric  case,
we may consider the IYIT model \cite{Izawa:1996pk,Intriligator:1996pu},
where there are four superfields of $\Psi_j~(i,j=1,2,3,4)$ with fundamental representation 
under $SU(2)$ gauge theory  
and six singlet superfields $\Phi^{ij} = -\Phi^{ji}$. The low energy superpotential becomes
\bea
W = {\vartheta}_3 (v,q ) \Phi^{ij} [\Psi_i \Psi_j]
\eea
with ${\rm Pf}([\Psi_i \Psi_j]) = \Lambda_{SU(2)}^4$.
Here, $[\Psi_i^c \Psi_j]$ is a composite operator  and $\Lambda_{SU(2)}$ is dynamically generated scale 
under the $SU(2)$ theory. Thus, scalar potential is given by the F-component of $\Phi$:
\bea
V = \Lambda_{SU(2)}^4  |{\vartheta}_3 (v,q )| ^2 +{\cal O}(\Phi).
\eea
Here, we have assumed that $\Phi =0$ is the minimum owing to the K\"ahler potential 
$K = |\Phi|^2 - c |\Phi|^4/\Lambda_{SU(2)}^2$, in which $c= {\cal O}( 10^{-2})$ is 
obtained after integrating out heavy modes with the masses of order $\Lambda_{SU(2)}$.
Alternatively, one may  consider the trilinear scalar term of the Yukawa interaction, assuming
the scalars develop non-zero VEVs by some dynamics. In this case the scalar potential is
approximately given by $V \sim m_{3/2} {\vartheta}_3 (v,q ) \la \Phi \ra^3$.

\subsection{Supergravity models with moduli stabilization}

In a more realistic stringy set-up one has to consider both real and imaginary components of $v$. 
In order to realize an effective single-field inflation, one needs to ensure that the other degrees of freedom
including the imaginary component of $v$ are stabilized with a mass heavier than (or comparable to)
the Hubble parameter during inflation. 
To see this, let us consider a supergravity model in type IIB orientifold with flux compactification 
\cite{Kachru:2003aw,Grimm:2004uq,Balasubramanian:2005zx,Conlon:2005ki,Choi:2006bh,Haack:2006cy,Camara:2003ku,Camara:2004jj}.

The following is the simplified effective action 
\bea
K &=& K(S+\bar{S};T+\bar{T}) - 3\log[-i(\tau -\bar \tau)] \non\\
&&+\sum_{\alpha} 
F_\alpha^2 |v_\alpha -\bar{v}_\alpha|^2 
+\cdots , \\
W &=& \int G \wedge \Omega (S, \tau, v_{\alpha}) + 
A(S, \tau, v_{\alpha},\Psi) e^{-aT}  \non\\ &&+ 
W_{\rm hidden~Yukawa}(v_\alpha,\tau,\Psi)
+ \cdots,
\label{effsugra}
\eea
where $S$ is a dilaton-axion, $T$ is a K\"ahler modulus, $\tau$ is a complex structure modulus,
and $v_\alpha$ is a open string modulus.
$F_\alpha$ is a decay constant for open string modulus, e.g.,
$F_W^2 = 3/(T+\bar T)$ and $F_P^2 =1/(S+\bar{S})$ for  D7 Wilson line moduli and D7 position moduli, respectively \cite{Jockers:2004yj},
and they are expected to be comparable to the string/Planck scale. We shall return to the issue of the size of the decay constant later.
$G$ is the three form flux,
$\Omega$ is holomorphic three form and the second term in the superpotential arises from
instantons or gaugino condensations. We have included trilinear superpotential $W_{\rm hidden~Yukawa}$
between matter-like fields $\Psi_i$ in the hidden sector.
In the K\"ahler potntial, the ellipsis include the terms for $\Psi_i$ and 
SUSY breaking sector and the quantum corrections \cite{Berg:2005ja}
in addtion to the Standard Model (SM) fields. The ellipsis in the superpotential 
contains contributions from 
SUSY-breaking and the SM sector.

With the above effective action,  $\tau$ and $S$ can be stabilized by a choice of three form fluxes $G$: 
$\int G \wedge \Omega \supset \tau^2 + S\tau $.
We may obtain a light open string moduli  $v$, a certain combination of $v_\alpha$, if
\bea
\frac{\partial}{\partial v} \int G \wedge \Omega (S, \tau, v_{\alpha}) = 0.
\eea
The other degrees of freedom in $\{v_\alpha\}$ are assumed to be stabilized with a heavy mass by
 $\int G \wedge \Omega \supset v_{\alpha}^2$, and therefore can be integrated out in the low energy.
For instance,  the Wilson line moduli are not included in flux-induced superpotential 
because of the internal gauge symmetry \cite{Camara:2004jj}. 
Alternatively, only a part of the position moduli on D-branes may be fixed by 
a selection of the closed string fluxes, magnetic (worldvolume) fluxes and geometry 
through the tadpole condition~\cite{Collinucci:2008pf}.
({Similar argument can be applied also to the complex structure moduli \cite{Abe:2006xi, Abe:2014xja}}
and to type IIA moduli stabilization without metric fluxes \cite{DeWolfe:2005uu}.) 
Finally,  through instantons or gaugino condensations on the stack of branes, 
the K\"ahler moduli $T$ can be stabilized taking account of the SUSY breaking.
The SUSY breaking also stabilizes the remaining moduli and 
produces the scalar potential of the Jacobi theta function 
through a gaugino mass or a trilinear scalar term. 
Here, we expect that ${\rm Im}(v)$ obtains the gravitino mass $m_{3/2} \propto \int G \wedge \Omega$
via the SUSY-breaking effect because the K\"ahler potential
depends only on the imaginary part of $v$ as in Eq.~(\ref{effsugra}) at the tree level.
Thus,  ${\rm Re}(v) \equiv \phi/2\pi F$  will be the lightest degrees of freedom,  a good candidate for the inflaton.

In the following we consider various ways to generate the inflaton potential, which depends on the Jacobi theta functions,
$\vartheta_i(v, q)$. 
We will discuss also an issue of one-loop breakdown of the shift symmetry along ${\rm Re}(v)$. 

\subsubsection*{Case 1: Gauge coupling depending on the theta function}

Let us first focus on the dependence on $v$ through a gauge coupling in the strongly coupled hidden sector.
This $v$-dependence appears in the non-perturbative term as we shall see below.

On the stack of $N$ D7-branes,
the holomorphic gauge coupling at one-loop level contains the Jacobi theta function
in the presence of e.g.  a stack of $n$ D3-branes \cite{Berg:2004ek}:
\bea
f_{\rm one-loop} \supset - \frac{n}{2\pi} \log(\vartheta_1(v, q)),
\eea
where $q=e^{i \pi \tau}$ is given by a torus complex structure modulus $\tau$
and we have assumed that the gauge coupling depends on $v$, and
the tree level coupling is given by $f_{\rm tree} = T$.
This is regarded as a backreaction from $n$ D3-branes to the worldvolume of $N$ 
D7-branes \cite{Baumann:2006th}. 
In this case $v$ denotes the position of D3-branes on the torus.
When the $N$ D7-branes undergo  gaugino condensation\footnote{
Similar type of the superpotential has been observed through D-instanton contributions
in Ref.~\cite{Grimm:2007xm,Grimm:2007hs} 
with $SL(2,{\mathbb Z})$ duality in type IIB supergravity compactified 
on Calabi-Yau orientifold with odd parity K\"ahler moduli ${\cal G}$:
For instance,
$W \sim \theta_m ({\cal G},iS) \eta^{-10}(iS)$ may be generated by D-instantons. 
Here, $\theta_m ({\cal G},iS) = \sum_{n \in  L+r} e^{-S n{^2}/2}e^{i m {\cal G}n}$ is
the theta function of weight $s/2$ with index $m$.
$L$ is some positive definite rational lattice of dimension $s$, 
and $r$ is some vector which admits an expansion in a basis of 
$L$ with rational coefficients. 
Then, we have $\partial_{\cal G} \int G \wedge \Omega = 0$, 
hence ${\rm Re}({\cal G})$ can be an inflaton candidate if exists, 
because there will be just ${\rm Im}({\cal G})$ dependence in the K\"ahler potential
at a perturbative level. 
This looks similar to the superpotential discussed in inflation model
with Dedekind eta function. 
}, the second non-perturbative term in
Eq.~(\ref{effsugra}) becomes
\bea
W_{\rm non-pert}
&\propto & 
\exp\bigg[-\frac{2\pi}{N}\bigg(T- \frac{n}{2\pi} \log(\vartheta_1(v, q) \bigg) \bigg]  \\
&\equiv & \Lambda_{SU(N)}^3 \, \bigg(\vartheta_1(v,q)\bigg)^{\frac{n}{N}}. 
\eea
Here, $\Lambda_{SU(N)}$ 
is the dynamically generated scale under the pure $SU(N)$ super Yang-Mills theory.
Now we have the gaugino mass term $m_{3/2} \lambda \lambda$ in the $SU(N)$ sector,
the scalar potential is given by
\bea
V \sim m_{3/2} \Lambda^3_{SU(N)} \, \bigg(\vartheta_1(v,q)\bigg)^{\frac{n}{N}} 
+ {\cal O}(\Lambda_{SU(N)}^6),
\eea
where we neglected the presence of matter $\Psi_i$ and higher order terms in $\Lambda_{SU(N)}$.
The inflaton is identified with the real component of $v$, the position of D3-branes: ${\rm Re}(v) \equiv \phi/2\pi F$. 
Thus, the inflaton potential depending on $\vartheta_1$ arises from the interactions between $n$ 
D3 branes and $N$ D7 branes through gaugino condensation in this case.

If the inflation is driven by the above inflaton potential, the Hubble scale during the inflation will be given by
$H_{\rm inf}^2 \sim m_{3/2} \Lambda^3_{SU(N)}$. 
In the KKLT case \cite{Kachru:2003aw}, 
we expect  $\Lambda_{SU(N)}^3 \lesssim m_{3/2} \lesssim \Lambda_{SU(N)}$, and so, $H_{\rm inf} \lesssim m_{3/2}$.
Therefore we may be able to avoid the decompactification problem~\cite{Kallosh:2004yh}. 
The potential with the double gaugino condensations can ameliorate this problem. 
(See also discussions in Ref.~\cite{Hayashi:2014aua}.)
The scalar potential induced by a similar mechanism was used to drive slow-roll inflation
in the context of the warped brane inflation \cite{Kachru:2003sx,Baumann:2006th,Baumann:2007np,Baumann:2007ah,Chen:2009nk},
where  inflation is driven by moving D3-branes in the warped geometry with
anti D3-branes at the tip. On the other hand, here we consider the inflation with a D-brane modulus in the bulk extra dimension,
and the inflaton potential arises from the interactions between D3-branes and D7-branes.
We may use anti D3-branes at the tip of a warped throat to uplift the AdS vacuum to the dS one with a small and positive
cosmological constant. The interaction between branes and anti branes can be neglected if
 it is a sufficiently strongly warped throat. Alternatively, we may consider
a field-theoretic spontaneous SUSY breaking for uplifting~\cite{Lebedev:2006qq,Dudas:2006gr,Abe:2006xp,Kallosh:2006dv}.
In either case the uplifting sector is required to preserve an approximate 
shift symmetry along inflaton direction \cite{Kachru:2003sx} for successful slow roll inflation.

Lastly, we note that
there are $v$-dependent one-loop corrections in the K\"ahler potential  \cite{Berg:2005ja},
which  induce the mass term for the inflaton $V \sim c \, m_{3/2}^2 \phi^2$ with an
expectation of $c = {\cal O}(10^{-2})$ or so. For successful slow-roll inflation, such corrections must be sufficiently small, 
$c \lesssim 10^{-2} (\Lambda_{SU(N)}^3/m_{3/2})$.

\subsubsection*{Case 2: Yukawa coupling depending on the theta function}

As noted already,  Yukawa couplings are given by the Jacobi theta function
in the toroidal compactification of extra dimension with magnetic fluxes in the string theory.
The superpotential is given by
\bea
W = {\vartheta}_3 (v,q ) \Psi_i \Psi_j \Psi_k  ,
\eea
where $\Psi_{i,j,k}$ are hidden matter-like superfields localized by the magnetic fluxes on the extra dimension,
$v$ is a open string modulus and $q$ is a torus complex structure moduli.
All the moduli other than $v$ are assumed to be stabilized with a heavy mass.
Taking account of the SUSY breaking effects, there will be a SUSY-breaking trilinear scalar term:
$V_{\rm SUSY-breaking} = m_{3/2} \vartheta_3 (v, q)  \Psi_i \Psi_j \Psi_k + c.c..$
(Here, note that  $\Psi_{i,j,k}$ are scalar component of the respective superfields.)
The scalars can develop non-zero VEVs, for instance, via D-term potentials involving with 
Fayet-Iliopoulos term $\xi$ generated by the magnetic fluxes and stabilized $T$: 
$D \propto \xi + \sum_i q_i |\Psi_i|^2$, where $q_i$ is an $U(1)$ charge of $\Psi_i$.
{See, e.g., Refs.\cite{Choi:2006bh,Dudas:2007nz,Dudas:2008qf,Li:2014owa,Li:2014xna,Li:2014lpa} 
for moduli stabilization with D-term potential.}
Then the potential becomes:
\bea
V \sim  a \vartheta_3 (v, q) + b \bigg(\vartheta_3 (v, q)\bigg)^2 + {\cal O}(\Psi^5) .
\eea
Here, ${\rm Re}(v) \equiv \phi/2\pi F$ is a candidate for the inflaton,
$a ={\cal O}(m_{3/2} \Psi^3)$, $b ={\cal O}(\Psi^4)$, and
we have added SUSY F-term contribution $V_{\rm SUSY} \propto  |\partial_{\Psi} W|^2$,
and neglected supergravity contributions of ${\cal O}(\Psi^5)$, which contain derivatives of 
the theta function.
Whether decompactification occurs depends the VEVs of $\{\Psi_i\}$, and $a,b \lesssim m_{3/2}^2$
is requird.
As discussed in the previous example, we  need to suppress one-loop corrections to the inflaton mass
for successful slow-roll inflation: $c \lesssim 10^{-2}a/m_{3/2}^2$ or
$c \lesssim 10^{-2}b/m_{3/2}^2$. 
Here $c$ is a coefficient of the one-loop induced inflaton mass term
normalized by the gravitino mass squared.

\subsubsection*{Case 3: Yukawa coupling depending on the theta function and strong dynamics}

Now we  consider a  scenario similar to the case 2, but $\Psi_i$ are stabilized by a strong dynamics 
as in Refs.~\cite{Choi:2006bh,Intriligator:1995au}:
\bea
W_{\rm non-pert} &=& (N_c -N_f)\bigg( 
\frac{A e^{-2 \pi T}}{{\rm det}(\Psi^c \Psi)}
\bigg)^{\frac{1}{N_c-N_f}}, \\
W_{\rm hidden~Yukawa} &=& {\vartheta}_3 (v,q ) \Phi \Psi^c \Psi ,
\eea
where $A$ includes $\vartheta_1$-dependent term in the non-perturbative effect as discussed in
the case 1,
$\Psi^c + \Psi$ are vector-like superfields charged under  $SU(N_c)$ gauge group 
with the number of $N_f (< N_c)$ flavors, and
$\Phi$ is a singlet superfields. 
Note that there are no open string 
moduli (adjoint modes) in the strong coupling brane to realize an asymptotically freedom.
Here, let us consider three stacks of D-branes labeled by $(\alpha, \beta, \gamma)$ for simplicity;
\bea
v =  I_{\alpha} v_\alpha  + I_{\beta} v_\beta + I_{\gamma} v_\gamma
\eea
is a linear combination of open string moduli $(v_{\alpha},v_{\beta},v_{\gamma})$ 
on the respective branes \cite{Cremades:2003qj,Cremades:2004wa}, and
${\rm Re}(v) \equiv \phi/2\pi F$ is the  inflaton.
Here, $I_{\alpha} = 
{\rm Ind}(D)_{\beta \gamma}$, $I_{\beta} = 
{\rm Ind}(D)_{\gamma \alpha}$, and $I_{\gamma} = 
{\rm Ind}(D)_{\alpha \beta}$, where ${\rm Ind}(D)_{\alpha \beta} 
\in {\mathbb Z}$ is the index of Dirac operator, which depends on magnetic flux
between the $\alpha$- and $\beta$-th D-brane and so on\footnote{
Dirac operator is given by $D_{\alpha \beta} = \partial + iA_{\alpha} - iA_{\beta}$,
where $A_{\alpha}$ denotes the background of gauge potential on the $\alpha$-th D-brane and so on. 
}. 
Hence $I_{\alpha}=-I_{\beta} = N_f$ and $I_{\gamma}$ counts the number of singlet $\Phi$.
Note that $v_{\gamma} \equiv 0$ when there exists 
the strongly coupled gauge theory on the $\gamma$-th $D$-brane,
because we have assumed there is no open string moduli on $\gamma$-th $D$-brane.
Note that $q$ is given by
$q = \exp(i \pi \tau |I_{\alpha}I_{\beta}I_{\gamma}|)$,
where $\tau$ is a complex structure moduli on the torus.

Here, we shall consider stabilization of matter-like fields. 
$\Phi$ can be fixed by D-term potential in the presence of Fayet-Iliopoulos term 
as discussed in the case 2:\footnote{
$D$ will develop the VEV as $D \sim m_{3/2}^2$.
}
\bea
D \propto \xi - |\Phi|^2 \sim 0.
\eea
This implies that $\Phi$ multiplet is eaten by the $U(1)$ gauge multiplet on which we are focusing. 
Here, we have used the canonical wave function for simplicity and $\xi$ is comparable to the cutoff scale.
Thus $\langle \Phi \rangle$ plays a role of heavy mass against vector-like pairs, and
they condense as $\langle \Psi^c \Psi \rangle =
 A^{\frac{1}{N_c}} e^{-\frac{2 \pi}{N_c} T} \big(\vartheta_3(v,q)\big)^{\frac{N_f}{N_c}-1}
\langle \Phi \rangle^{\frac{N_f}{N_c}-1}$.
We integrate out them and obtains low energy effective potential
\bea
W_{\rm eff} &=& N_c A^{\frac{1}{N_c}} e^{-\frac{2 \pi}{N_c} T} \big(\vartheta_3(v,q)\big)^{\frac{N_f}{N_c}}
\langle \Phi \rangle^{\frac{N_f}{N_c}}  \\
&\equiv & \Lambda_{SU(N_c)}^3 A^{\frac{1}{N_c}}\big(\vartheta_3(v,q)\big)^{\frac{N_f}{N_c}} .
\eea
Here, $\Lambda_{SU(N_c)}$ is defined as 
the dynamically generated scale in the pure $SU(N_c)$ Yang-Mills theory
in the low energy.
Thus, low energy potential can be rewritten as
$V \sim m_{3/2}\Lambda_{SU(N_c)}^3 \big(\vartheta_3(v,q)\big)^{\frac{N_f}{N_c}} $
through the gaugino mass and the trilinear scalar term between $\Phi$ and $\Psi^c + \Psi$,
if $A$ does not depend on $v$. We expect, however, that two stacks of flavor branes can give an effect 
on the worldvolume of strong coupling brane, hence we will find
$A \sim [\vartheta_1(v_{\alpha},\tilde{q} )]^{n_{\alpha}}
[\vartheta_1(v_{\beta},\tilde{q} )]^{n_{\beta}}$, where $\tilde q = e^{i \pi \tau}$, $n_{\alpha}$ and $n_{\beta}$
denote the number of $\alpha$-th and $\beta$-th D-branes for flavors: Two $\vartheta_1$ terms come from backreaction 
of $\alpha$-th and $\beta$-th D-branes to the $\gamma$-th brane. 
Thus the scalar potential is
\bea
V &\sim& m_{3/2}\Lambda_{SU(N_c)}^3 \bigg[\vartheta_1\bigg(\frac{v}{2N_f},\tilde{q} \bigg)\bigg]^{\frac{2}{N_c}}
\bigg[\vartheta_3(v,q)\bigg]^{\frac{N_f}{N_c}} \non\\ &&
+ {\cal O}(\Lambda_{SU(N_c)}^6) .
\eea
Here we have assumed that $v_{\alpha} + v_{\beta}$ is stabilized by closed string fluxes around 
the origin,
$v/N_f = v_{\alpha} - v_{\beta}$, $\vartheta_1(-v,q) = - \vartheta_1(v,q)$, and 
we took $n_{\alpha}=n_{\beta}=1$.
This potential is similar to the case 1, but now it is accompanied with the $\vartheta_3$-dependent term.
Note that the decay constant is enhanced by $2N_f$ in the $\vartheta_1$ term
compared to that in the  $\vartheta_3$ term.
.
%

{

\subsection{Potential with a gauge coupling depending on the Dedekind eta function}

Now let us consider inflation models involving the Dedekind eta function, in which
a complex structure modulus $\tau$ is an inflaton candidate \cite{Abe:2014xja}
and a running spectral index is generated naturally.
See also Refs.\cite{Font:1990nt,Font:1990gx} and \cite{Abe:2014pwa} in Heterotic cases.
We choose closed sting fluxes such that $\partial_{\tau} \int G \wedge \Omega = 0$
in Eq.~(\ref{effsugra}) with assumption that all open string moduli are stabilized.
It is known that the a holomorphic gauge coupling includes the Dedekind eta function
\cite{Berg:2004ek,Blumenhagen:2006ci}:
\bea
f_{\rm one-loop} \supset - \frac{b}{2\pi}\log (\eta(\tau)).
\eea
Here this is obtained after integrating out open string modes stretched among D-branes
which are propagating on a part of the bulk
extra dimension and hence $b$ denotes 1-loop coefficient of beta function for the gauge coupling
in ${\cal N}=2$ SUSY sector.
We expect that, as in the case 1, 
the gauge coupling can depend on $\vartheta_1(\langle v\rangle ,q)$, 
where $q=e^{i \pi \tau}$, when there is another stack of $n$ D-branes.
Including these contributions to the gauge coupling with $f_{\rm tree} = T$, 
the gaugino condensation in the pure $SU(N)$ theory is given as
\bea
\nonumber
W_{\rm non-pert}
&\propto & 
\exp\bigg[-\frac{2\pi}{N}\bigg(
T-\frac{b}{2\pi}\log (\eta(\tau)) 
\bigg) \bigg]
e^{ \frac{n}{N} \log(\vartheta_1(\langle v \rangle, q))}  
\\
&\equiv & \Lambda_{SU(N)}^3 \, \big[ \eta (\tau)\big]^{\frac{b}{N}}
\big(\vartheta_1(\langle v \rangle, q) \big)^{\frac{n}{N}}. 
\eea
Here, $\Lambda_{SU(N)}$ is the dynamically generated scale in the $SU(N)$.
Though the gaugino mass term depending on the gravitino mass, 
the scalar potential includes
\bea
\nonumber
V &\sim& 
m_{3/2} \Lambda_{SU(N)}^3 \big[ \eta (\tau)\big]^{\frac{b}{N}}
\big(\vartheta_1(\langle v \rangle, q) \big)^{\frac{n}{N}}
 + c.c. 
\\
&& 
~~~~~~~~~~~~~~~~~~~~~~~~~~~~~~~~
+ {\cal O}(\Lambda_{SU(N)}^6).
\eea
In Refs.~\cite{Font:1990nt,Font:1990gx} on Heterotic otbifolds,  
$b/N = -2$ can be obtained with a requirement of the modular invariance along $\tau$ direction.
If $\tau $ is a kind of overall modulus, we find $b/N = -6$.

}

\subsection{Super-Planckian decay constant}
The decay constant for the inflaton is expected to be of order the string/Planck scale.
It is, however, possible to realize a decay constant larger than the Planck scale
by the alignment mechanism in the presence of multiple light moduli \cite{Kim:2004rp}.
For instance, when there are two  Yukawa couplings
and two light brane moduli $v_1$ and $v_2$, i.e.,
$ \partial_{v_{1,2}}  \int G \wedge \Omega = 0$ and
$ W_{\rm hidden~Yukawa} 
= \vartheta_3 (I_1v_1 + I_2v_2,q)\Psi^3 + \vartheta_3 (J_1v_1 + J_2 v_2, q)\Psi'^3,$
an enhanced decay constant can be obtained with
$|J_1 -J_2|/I < 1$, $I_1 = I_2 \equiv I$ and $\langle \Psi' \rangle < \langle \Psi\rangle $.
The inflaton becomes ${\rm Re}(v_1 - v_2)$ 
and the decay constant is enhanced by a factor of $I/|J_1-J_2|$.
Indeed, we have seen the enhancement of the decay constant by $2N_f$ in the scalar potential 
in the above case 3.

\section{Discussion and Conclusions}
The Universe must be reheated by the inflaton decay some time after inflation. 
The inflaton appears in the gauge and Yukawa couplings, and so, it 
may decay into  the SM gauge bosons (or gauginios if kinematically accessible) through the gauge couplings
or into  the SM matter and the Higgs or right-handed neutrinos and $B-L$ Higgs through Yukawa couplings.  
For instance, the following operators will be relevant to the reheating process:
\bea
\frac{\phi}{F_v} (F_{\mu \nu})^2,~~~
\frac{\phi}{F_v} y_t \bar{t}t H,~~~
\frac{\phi}{F_v} y_N \varphi_{B-L} NN.
\eea
Here, we have taken $\langle \phi \rangle \neq 0$ because the modulus develops a non-zero VEV in general, and 
the decay constant $F_v$ is comparable to the string scale, which is not necessarily equal to
the decay constant in the inflaton potential. $F_{\mu \nu}$, $t$, $H$, $\varphi_{B-L}$ and $N$ 
denote the SM gauge bosons, top quark, $B-L$ Higgs and right-handed neutrino respectively.
The thermal leptogenesis is viable if the reheating temperature is enough high as $10^9$ GeV
\cite{Fukugita:1986hr}. 
The last operator will be useful to implement non-thermal leptogenesis,
even if the reheating temperature becomes lower \cite{Asaka:1999yd,Asaka:1999jb}.

\vspace{3mm}
In this paper we have studied an extension of the natural inflation, where the inflaton potential
has a periodic property. We have proposed elliptic inflation where the inflaton 
potential is based on the elliptic functions and related functions. Focusing on
the elliptic and theta inflation, we have shown that the predicted
$n_s$ and $r$ interpolate from natural inflation to the exponential inflation including $R^2$-inflation,
where $n_s$ asymptotes to $n_s \sim 1-2/N \simeq 0.967$ for $N=60$.
We have also studied the inflation model with the Dedekind eta function and shown
that a sizable running of the spectral index can be generated owing to the small modulations
on the inflaton potential. For a certain value of the parameter, it is possible to realize $dn_s/d\ln k \sim -0.01$ 
which is more or less constant over the CMB  scales.

Periodicity often appears  in the field theories and the string 
theories through compactifications.
Interestingly, such periodic backgrounds restrict the allowed form of the potential,
and as a result, the inflaton potential takes a specific form rather than simple cosine function(s).
For instance, toroidal background forces infinite series of cosine functions to add up to give the Jacobi theta functions
(cf. Eqs.~(\ref{theta0}) - (\ref{theta3})).
Indeed,  the Jacobi theta functions are known to appear in gauge
and Yukawa couplings in toroidal compactifications, where the variables are open string and complex
structure moduli. We have explored various ways to generate the scalar potential which depends
on the Jacobi theta functions. The inflaton is identified with the real component of the lightest open string modulus,
and we have discussed the stabilization of the other moduli fields based on a supergravity model in type IIB orientifold 
with flux compactification.

%
%
%

\acknowledgments
The authors would like to thank to Tatsuo Kobayashi for discussions on Yukawa interactions in string models.
This work was supported by  JSPS Grant-in-Aid for
Young Scientists (B) (No.24740135 [FT] and No. 25800169 [TH]), 
Scientific Research (A) (No.26247042 [TH, FT]), Scientific Research (B) (No.26287039 [FT]), 
 the Grant-in-Aid for Scientific Research on Innovative Areas (No.23104008 [FT]),  and
Inoue Foundation for Science [FT].  This work was also
supported by World Premier International Center Initiative (WPI Program), MEXT, Japan [FT].

\bibliography{JEFinflation}

\begin{thebibliography}{96}%
\makeatletter
\providecommand \@ifxundefined [1]{%
 \@ifx{#1\undefined}
}%
\providecommand \@ifnum [1]{%
 \ifnum #1\expandafter \@firstoftwo
 \else \expandafter \@secondoftwo
 \fi
}%
\providecommand \@ifx [1]{%
 \ifx #1\expandafter \@firstoftwo
 \else \expandafter \@secondoftwo
 \fi
}%
\providecommand \natexlab [1]{#1}%
\providecommand \enquote  [1]{``#1''}%
\providecommand \bibnamefont  [1]{#1}%
\providecommand \bibfnamefont [1]{#1}%
\providecommand \citenamefont [1]{#1}%
\providecommand \href@noop [0]{\@secondoftwo}%
\providecommand \href [0]{\begingroup \@sanitize@url \@href}%
\providecommand \@href[1]{\@@startlink{#1}\@@href}%
\providecommand \@@href[1]{\endgroup#1\@@endlink}%
\providecommand \@sanitize@url [0]{\catcode `\\12\catcode `\$12\catcode
  `\&12\catcode `\#12\catcode `\^12\catcode `\_12\catcode `\%12\relax}%
\providecommand \@@startlink[1]{}%
\providecommand \@@endlink[0]{}%
\providecommand \url  [0]{\begingroup\@sanitize@url \@url }%
\providecommand \@url [1]{\endgroup\@href {#1}{\urlprefix }}%
\providecommand \urlprefix  [0]{URL }%
\providecommand \Eprint [0]{\href }%
\providecommand \doibase [0]{http://dx.doi.org/}%
\providecommand \selectlanguage [0]{\@gobble}%
\providecommand \bibinfo  [0]{\@secondoftwo}%
\providecommand \bibfield  [0]{\@secondoftwo}%
\providecommand \translation [1]{[#1]}%
\providecommand \BibitemOpen [0]{}%
\providecommand \bibitemStop [0]{}%
\providecommand \bibitemNoStop [0]{.\EOS\space}%
\providecommand \EOS [0]{\spacefactor3000\relax}%
\providecommand \BibitemShut  [1]{\csname bibitem#1\endcsname}%
\let\auto@bib@innerbib\@empty
\bibitem [{\citenamefont {Guth}(1981)}]{Guth:1980zm}%
  \BibitemOpen
  \bibfield  {author} {\bibinfo {author} {\bibfnamefont {A.~H.}\ \bibnamefont
  {Guth}},\ }\href {\doibase 10.1103/PhysRevD.23.347} {\bibfield  {journal}
  {\bibinfo  {journal} {Phys.Rev.}\ }\textbf {\bibinfo {volume} {D23}},\
  \bibinfo {pages} {347} (\bibinfo {year} {1981})}\BibitemShut {NoStop}%
\bibitem [{\citenamefont {Sato}(1981)}]{Sato:1980yn}%
  \BibitemOpen
  \bibfield  {author} {\bibinfo {author} {\bibfnamefont {K.}~\bibnamefont
  {Sato}},\ }\href@noop {} {\bibfield  {journal} {\bibinfo  {journal}
  {Mon.Not.Roy.Astron.Soc.}\ }\textbf {\bibinfo {volume} {195}},\ \bibinfo
  {pages} {467} (\bibinfo {year} {1981})}\BibitemShut {NoStop}%
\bibitem [{\citenamefont {Starobinsky}(1980)}]{Starobinsky:1980te}%
  \BibitemOpen
  \bibfield  {author} {\bibinfo {author} {\bibfnamefont {A.~A.}\ \bibnamefont
  {Starobinsky}},\ }\href {\doibase 10.1016/0370-2693(80)90670-X} {\bibfield
  {journal} {\bibinfo  {journal} {Phys.Lett.}\ }\textbf {\bibinfo {volume}
  {B91}},\ \bibinfo {pages} {99} (\bibinfo {year} {1980})}\BibitemShut
  {NoStop}%
\bibitem [{\citenamefont {Brout}\ \emph {et~al.}(1978)\citenamefont {Brout},
  \citenamefont {Englert},\ and\ \citenamefont {Gunzig}}]{Brout:1977ix}%
  \BibitemOpen
  \bibfield  {author} {\bibinfo {author} {\bibfnamefont {R.}~\bibnamefont
  {Brout}}, \bibinfo {author} {\bibfnamefont {F.}~\bibnamefont {Englert}}, \
  and\ \bibinfo {author} {\bibfnamefont {E.}~\bibnamefont {Gunzig}},\ }\href
  {\doibase 10.1016/0003-4916(78)90176-8} {\bibfield  {journal} {\bibinfo
  {journal} {Annals Phys.}\ }\textbf {\bibinfo {volume} {115}},\ \bibinfo
  {pages} {78} (\bibinfo {year} {1978})}\BibitemShut {NoStop}%
\bibitem [{\citenamefont {Kazanas}(1980)}]{Kazanas:1980tx}%
  \BibitemOpen
  \bibfield  {author} {\bibinfo {author} {\bibfnamefont {D.}~\bibnamefont
  {Kazanas}},\ }\href {\doibase 10.1086/183361} {\bibfield  {journal} {\bibinfo
   {journal} {Astrophys.J.}\ }\textbf {\bibinfo {volume} {241}},\ \bibinfo
  {pages} {L59} (\bibinfo {year} {1980})}\BibitemShut {NoStop}%
\bibitem [{\citenamefont {Linde}(1982)}]{Linde:1981mu}%
  \BibitemOpen
  \bibfield  {author} {\bibinfo {author} {\bibfnamefont {A.~D.}\ \bibnamefont
  {Linde}},\ }\href {\doibase 10.1016/0370-2693(82)91219-9} {\bibfield
  {journal} {\bibinfo  {journal} {Phys.Lett.}\ }\textbf {\bibinfo {volume}
  {B108}},\ \bibinfo {pages} {389} (\bibinfo {year} {1982})}\BibitemShut
  {NoStop}%
\bibitem [{\citenamefont {Albrecht}\ and\ \citenamefont
  {Steinhardt}(1982)}]{Albrecht:1982wi}%
  \BibitemOpen
  \bibfield  {author} {\bibinfo {author} {\bibfnamefont {A.}~\bibnamefont
  {Albrecht}}\ and\ \bibinfo {author} {\bibfnamefont {P.~J.}\ \bibnamefont
  {Steinhardt}},\ }\href {\doibase 10.1103/PhysRevLett.48.1220} {\bibfield
  {journal} {\bibinfo  {journal} {Phys.Rev.Lett.}\ }\textbf {\bibinfo {volume}
  {48}},\ \bibinfo {pages} {1220} (\bibinfo {year} {1982})}\BibitemShut
  {NoStop}%
\bibitem [{\citenamefont {Starobinsky}(1979)}]{Starobinsky:1979ty}%
  \BibitemOpen
  \bibfield  {author} {\bibinfo {author} {\bibfnamefont {A.~A.}\ \bibnamefont
  {Starobinsky}},\ }\href@noop {} {\bibfield  {journal} {\bibinfo  {journal}
  {JETP Lett.}\ }\textbf {\bibinfo {volume} {30}},\ \bibinfo {pages} {682}
  (\bibinfo {year} {1979})}\BibitemShut {NoStop}%
\bibitem [{\citenamefont {Linde}(1983)}]{Linde:1983gd}%
  \BibitemOpen
  \bibfield  {author} {\bibinfo {author} {\bibfnamefont {A.~D.}\ \bibnamefont
  {Linde}},\ }\href {\doibase 10.1016/0370-2693(83)90837-7} {\bibfield
  {journal} {\bibinfo  {journal} {Phys.Lett.}\ }\textbf {\bibinfo {volume}
  {B129}},\ \bibinfo {pages} {177} (\bibinfo {year} {1983})}\BibitemShut
  {NoStop}%
\bibitem [{\citenamefont {Freese}\ \emph {et~al.}(1990)\citenamefont {Freese},
  \citenamefont {Frieman},\ and\ \citenamefont {Olinto}}]{Freese:1990rb}%
  \BibitemOpen
  \bibfield  {author} {\bibinfo {author} {\bibfnamefont {K.}~\bibnamefont
  {Freese}}, \bibinfo {author} {\bibfnamefont {J.~A.}\ \bibnamefont {Frieman}},
  \ and\ \bibinfo {author} {\bibfnamefont {A.~V.}\ \bibnamefont {Olinto}},\
  }\href {\doibase 10.1103/PhysRevLett.65.3233} {\bibfield  {journal} {\bibinfo
   {journal} {Phys.Rev.Lett.}\ }\textbf {\bibinfo {volume} {65}},\ \bibinfo
  {pages} {3233} (\bibinfo {year} {1990})}\BibitemShut {NoStop}%
\bibitem [{\citenamefont {Ade}\ \emph {et~al.}(2013)\citenamefont {Ade} \emph
  {et~al.}}]{Ade:2013uln}%
  \BibitemOpen
  \bibfield  {author} {\bibinfo {author} {\bibfnamefont {P.}~\bibnamefont
  {Ade}} \emph {et~al.} (\bibinfo {collaboration} {Planck Collaboration}),\
  }\href@noop {} {\  (\bibinfo {year} {2013})},\ \Eprint
  {http://arxiv.org/abs/1303.5082} {arXiv:1303.5082 [astro-ph.CO]} \BibitemShut
  {NoStop}%
\bibitem [{\citenamefont {Ade}\ \emph {et~al.}(2014{\natexlab{a}})\citenamefont
  {Ade} \emph {et~al.}}]{Ade:2014xna}%
  \BibitemOpen
  \bibfield  {author} {\bibinfo {author} {\bibfnamefont {P.}~\bibnamefont
  {Ade}} \emph {et~al.} (\bibinfo {collaboration} {BICEP2 Collaboration}),\
  }\href {\doibase 10.1103/PhysRevLett.112.241101} {\bibfield  {journal}
  {\bibinfo  {journal} {Phys.Rev.Lett.}\ }\textbf {\bibinfo {volume} {112}},\
  \bibinfo {pages} {241101} (\bibinfo {year} {2014}{\natexlab{a}})},\ \Eprint
  {http://arxiv.org/abs/1403.3985} {arXiv:1403.3985 [astro-ph.CO]} \BibitemShut
  {NoStop}%
\bibitem [{\citenamefont {Adam}\ \emph {et~al.}(2014)\citenamefont {Adam} \emph
  {et~al.}}]{Adam:2014oea}%
  \BibitemOpen
  \bibfield  {author} {\bibinfo {author} {\bibfnamefont {R.}~\bibnamefont
  {Adam}} \emph {et~al.} (\bibinfo {collaboration} {Planck Collaboration}),\
  }\href@noop {} {\  (\bibinfo {year} {2014})},\ \Eprint
  {http://arxiv.org/abs/1409.5738} {arXiv:1409.5738 [astro-ph.CO]} \BibitemShut
  {NoStop}%
\bibitem [{\citenamefont {Bezrukov}\ and\ \citenamefont
  {Shaposhnikov}(2008)}]{Bezrukov:2007ep}%
  \BibitemOpen
  \bibfield  {author} {\bibinfo {author} {\bibfnamefont {F.~L.}\ \bibnamefont
  {Bezrukov}}\ and\ \bibinfo {author} {\bibfnamefont {M.}~\bibnamefont
  {Shaposhnikov}},\ }\href {\doibase 10.1016/j.physletb.2007.11.072} {\bibfield
   {journal} {\bibinfo  {journal} {Phys.Lett.}\ }\textbf {\bibinfo {volume}
  {B659}},\ \bibinfo {pages} {703} (\bibinfo {year} {2008})},\ \Eprint
  {http://arxiv.org/abs/0710.3755} {arXiv:0710.3755 [hep-th]} \BibitemShut
  {NoStop}%
\bibitem [{\citenamefont {Spokoiny}(1984)}]{Spokoiny:1984bd}%
  \BibitemOpen
  \bibfield  {author} {\bibinfo {author} {\bibfnamefont {B.}~\bibnamefont
  {Spokoiny}},\ }\href {\doibase 10.1016/0370-2693(84)90587-2} {\bibfield
  {journal} {\bibinfo  {journal} {Phys.Lett.}\ }\textbf {\bibinfo {volume}
  {B147}},\ \bibinfo {pages} {39} (\bibinfo {year} {1984})}\BibitemShut
  {NoStop}%
\bibitem [{\citenamefont {Lucchin}\ and\ \citenamefont
  {Matarrese}(1985)}]{Lucchin:1984yf}%
  \BibitemOpen
  \bibfield  {author} {\bibinfo {author} {\bibfnamefont {F.}~\bibnamefont
  {Lucchin}}\ and\ \bibinfo {author} {\bibfnamefont {S.}~\bibnamefont
  {Matarrese}},\ }\href {\doibase 10.1103/PhysRevD.32.1316} {\bibfield
  {journal} {\bibinfo  {journal} {Phys.Rev.}\ }\textbf {\bibinfo {volume}
  {D32}},\ \bibinfo {pages} {1316} (\bibinfo {year} {1985})}\BibitemShut
  {NoStop}%
\bibitem [{\citenamefont {Goncharov}\ and\ \citenamefont
  {Linde}(1984)}]{Goncharov:1983mw}%
  \BibitemOpen
  \bibfield  {author} {\bibinfo {author} {\bibfnamefont {A.}~\bibnamefont
  {Goncharov}}\ and\ \bibinfo {author} {\bibfnamefont {A.~D.}\ \bibnamefont
  {Linde}},\ }\href {\doibase 10.1016/0370-2693(84)90027-3} {\bibfield
  {journal} {\bibinfo  {journal} {Phys.Lett.}\ }\textbf {\bibinfo {volume}
  {B139}},\ \bibinfo {pages} {27} (\bibinfo {year} {1984})}\BibitemShut
  {NoStop}%
\bibitem [{\citenamefont {Salopek}\ \emph {et~al.}(1989)\citenamefont
  {Salopek}, \citenamefont {Bond},\ and\ \citenamefont
  {Bardeen}}]{Salopek:1988qh}%
  \BibitemOpen
  \bibfield  {author} {\bibinfo {author} {\bibfnamefont {D.}~\bibnamefont
  {Salopek}}, \bibinfo {author} {\bibfnamefont {J.}~\bibnamefont {Bond}}, \
  and\ \bibinfo {author} {\bibfnamefont {J.~M.}\ \bibnamefont {Bardeen}},\
  }\href {\doibase 10.1103/PhysRevD.40.1753} {\bibfield  {journal} {\bibinfo
  {journal} {Phys.Rev.}\ }\textbf {\bibinfo {volume} {D40}},\ \bibinfo {pages}
  {1753} (\bibinfo {year} {1989})}\BibitemShut {NoStop}%
\bibitem [{\citenamefont {Fakir}\ and\ \citenamefont
  {Unruh}(1990)}]{Fakir:1990eg}%
  \BibitemOpen
  \bibfield  {author} {\bibinfo {author} {\bibfnamefont {R.}~\bibnamefont
  {Fakir}}\ and\ \bibinfo {author} {\bibfnamefont {W.}~\bibnamefont {Unruh}},\
  }\href {\doibase 10.1103/PhysRevD.41.1783} {\bibfield  {journal} {\bibinfo
  {journal} {Phys.Rev.}\ }\textbf {\bibinfo {volume} {D41}},\ \bibinfo {pages}
  {1783} (\bibinfo {year} {1990})}\BibitemShut {NoStop}%
\bibitem [{\citenamefont {Stewart}(1995)}]{Stewart:1994ts}%
  \BibitemOpen
  \bibfield  {author} {\bibinfo {author} {\bibfnamefont {E.~D.}\ \bibnamefont
  {Stewart}},\ }\href {\doibase 10.1103/PhysRevD.51.6847} {\bibfield  {journal}
  {\bibinfo  {journal} {Phys.Rev.}\ }\textbf {\bibinfo {volume} {D51}},\
  \bibinfo {pages} {6847} (\bibinfo {year} {1995})},\ \Eprint
  {http://arxiv.org/abs/hep-ph/9405389} {arXiv:hep-ph/9405389 [hep-ph]}
  \BibitemShut {NoStop}%
\bibitem [{\citenamefont {Dvali}\ and\ \citenamefont
  {Tye}(1999)}]{Dvali:1998pa}%
  \BibitemOpen
  \bibfield  {author} {\bibinfo {author} {\bibfnamefont {G.}~\bibnamefont
  {Dvali}}\ and\ \bibinfo {author} {\bibfnamefont {S.~H.}\ \bibnamefont
  {Tye}},\ }\href {\doibase 10.1016/S0370-2693(99)00132-X} {\bibfield
  {journal} {\bibinfo  {journal} {Phys.Lett.}\ }\textbf {\bibinfo {volume}
  {B450}},\ \bibinfo {pages} {72} (\bibinfo {year} {1999})},\ \Eprint
  {http://arxiv.org/abs/hep-ph/9812483} {arXiv:hep-ph/9812483 [hep-ph]}
  \BibitemShut {NoStop}%
\bibitem [{\citenamefont {Cicoli}\ \emph {et~al.}(2009)\citenamefont {Cicoli},
  \citenamefont {Burgess},\ and\ \citenamefont {Quevedo}}]{Cicoli:2008gp}%
  \BibitemOpen
  \bibfield  {author} {\bibinfo {author} {\bibfnamefont {M.}~\bibnamefont
  {Cicoli}}, \bibinfo {author} {\bibfnamefont {C.}~\bibnamefont {Burgess}}, \
  and\ \bibinfo {author} {\bibfnamefont {F.}~\bibnamefont {Quevedo}},\ }\href
  {\doibase 10.1088/1475-7516/2009/03/013} {\bibfield  {journal} {\bibinfo
  {journal} {JCAP}\ }\textbf {\bibinfo {volume} {0903}},\ \bibinfo {pages}
  {013} (\bibinfo {year} {2009})},\ \Eprint {http://arxiv.org/abs/0808.0691}
  {arXiv:0808.0691 [hep-th]} \BibitemShut {NoStop}%
\bibitem [{\citenamefont {Nakayama}\ \emph
  {et~al.}(2013{\natexlab{a}})\citenamefont {Nakayama}, \citenamefont
  {Takahashi},\ and\ \citenamefont {Yanagida}}]{Nakayama:2013jka}%
  \BibitemOpen
  \bibfield  {author} {\bibinfo {author} {\bibfnamefont {K.}~\bibnamefont
  {Nakayama}}, \bibinfo {author} {\bibfnamefont {F.}~\bibnamefont {Takahashi}},
  \ and\ \bibinfo {author} {\bibfnamefont {T.~T.}\ \bibnamefont {Yanagida}},\
  }\href {\doibase 10.1016/j.physletb.2013.06.050} {\bibfield  {journal}
  {\bibinfo  {journal} {Phys.Lett.}\ }\textbf {\bibinfo {volume} {B725}},\
  \bibinfo {pages} {111} (\bibinfo {year} {2013}{\natexlab{a}})},\ \Eprint
  {http://arxiv.org/abs/1303.7315} {arXiv:1303.7315 [hep-ph]} \BibitemShut
  {NoStop}%
\bibitem [{\citenamefont {Nakayama}\ \emph
  {et~al.}(2013{\natexlab{b}})\citenamefont {Nakayama}, \citenamefont
  {Takahashi},\ and\ \citenamefont {Yanagida}}]{Nakayama:2013txa}%
  \BibitemOpen
  \bibfield  {author} {\bibinfo {author} {\bibfnamefont {K.}~\bibnamefont
  {Nakayama}}, \bibinfo {author} {\bibfnamefont {F.}~\bibnamefont {Takahashi}},
  \ and\ \bibinfo {author} {\bibfnamefont {T.~T.}\ \bibnamefont {Yanagida}},\
  }\href {\doibase 10.1088/1475-7516/2013/08/038} {\bibfield  {journal}
  {\bibinfo  {journal} {JCAP}\ }\textbf {\bibinfo {volume} {1308}},\ \bibinfo
  {pages} {038} (\bibinfo {year} {2013}{\natexlab{b}})},\ \Eprint
  {http://arxiv.org/abs/1305.5099} {arXiv:1305.5099} \BibitemShut {NoStop}%
\bibitem [{\citenamefont {Kallosh}\ \emph {et~al.}(2014)\citenamefont
  {Kallosh}, \citenamefont {Linde},\ and\ \citenamefont
  {Westphal}}]{Kallosh:2014xwa}%
  \BibitemOpen
  \bibfield  {author} {\bibinfo {author} {\bibfnamefont {R.}~\bibnamefont
  {Kallosh}}, \bibinfo {author} {\bibfnamefont {A.}~\bibnamefont {Linde}}, \
  and\ \bibinfo {author} {\bibfnamefont {A.}~\bibnamefont {Westphal}},\ }\href
  {\doibase 10.1103/PhysRevD.90.023534} {\bibfield  {journal} {\bibinfo
  {journal} {Phys.Rev.}\ }\textbf {\bibinfo {volume} {D90}},\ \bibinfo {pages}
  {023534} (\bibinfo {year} {2014})},\ \Eprint {http://arxiv.org/abs/1405.0270}
  {arXiv:1405.0270 [hep-th]} \BibitemShut {NoStop}%
\bibitem [{\citenamefont {Nakayama}\ \emph {et~al.}(2014)\citenamefont
  {Nakayama}, \citenamefont {Takahashi},\ and\ \citenamefont
  {Yanagida}}]{Nakayama:2014wpa}%
  \BibitemOpen
  \bibfield  {author} {\bibinfo {author} {\bibfnamefont {K.}~\bibnamefont
  {Nakayama}}, \bibinfo {author} {\bibfnamefont {F.}~\bibnamefont {Takahashi}},
  \ and\ \bibinfo {author} {\bibfnamefont {T.~T.}\ \bibnamefont {Yanagida}},\
  }\href {\doibase 10.1016/j.physletb.2014.08.043} {\bibfield  {journal}
  {\bibinfo  {journal} {Phys.Lett.}\ }\textbf {\bibinfo {volume} {B737}},\
  \bibinfo {pages} {151} (\bibinfo {year} {2014})},\ \Eprint
  {http://arxiv.org/abs/1407.7082} {arXiv:1407.7082 [hep-ph]} \BibitemShut
  {NoStop}%
\bibitem [{\citenamefont {Czerny}\ and\ \citenamefont
  {Takahashi}(2014)}]{Czerny:2014wza}%
  \BibitemOpen
  \bibfield  {author} {\bibinfo {author} {\bibfnamefont {M.}~\bibnamefont
  {Czerny}}\ and\ \bibinfo {author} {\bibfnamefont {F.}~\bibnamefont
  {Takahashi}},\ }\href {\doibase 10.1016/j.physletb.2014.04.039} {\bibfield
  {journal} {\bibinfo  {journal} {Phys.Lett.}\ }\textbf {\bibinfo {volume}
  {B733}},\ \bibinfo {pages} {241} (\bibinfo {year} {2014})},\ \Eprint
  {http://arxiv.org/abs/1401.5212} {arXiv:1401.5212 [hep-ph]} \BibitemShut
  {NoStop}%
\bibitem [{\citenamefont {Czerny}\ \emph
  {et~al.}(2014{\natexlab{a}})\citenamefont {Czerny}, \citenamefont {Higaki},\
  and\ \citenamefont {Takahashi}}]{Czerny:2014xja}%
  \BibitemOpen
  \bibfield  {author} {\bibinfo {author} {\bibfnamefont {M.}~\bibnamefont
  {Czerny}}, \bibinfo {author} {\bibfnamefont {T.}~\bibnamefont {Higaki}}, \
  and\ \bibinfo {author} {\bibfnamefont {F.}~\bibnamefont {Takahashi}},\ }\href
  {\doibase 10.1007/JHEP05(2014)144} {\bibfield  {journal} {\bibinfo  {journal}
  {JHEP}\ }\textbf {\bibinfo {volume} {1405}},\ \bibinfo {pages} {144}
  (\bibinfo {year} {2014}{\natexlab{a}})},\ \Eprint
  {http://arxiv.org/abs/1403.0410} {arXiv:1403.0410 [hep-ph]} \BibitemShut
  {NoStop}%
\bibitem [{\citenamefont {Czerny}\ \emph
  {et~al.}(2014{\natexlab{b}})\citenamefont {Czerny}, \citenamefont {Higaki},\
  and\ \citenamefont {Takahashi}}]{Czerny:2014qqa}%
  \BibitemOpen
  \bibfield  {author} {\bibinfo {author} {\bibfnamefont {M.}~\bibnamefont
  {Czerny}}, \bibinfo {author} {\bibfnamefont {T.}~\bibnamefont {Higaki}}, \
  and\ \bibinfo {author} {\bibfnamefont {F.}~\bibnamefont {Takahashi}},\ }\href
  {\doibase 10.1016/j.physletb.2014.05.041} {\bibfield  {journal} {\bibinfo
  {journal} {Phys.Lett.}\ }\textbf {\bibinfo {volume} {B734}},\ \bibinfo
  {pages} {167} (\bibinfo {year} {2014}{\natexlab{b}})},\ \Eprint
  {http://arxiv.org/abs/1403.5883} {arXiv:1403.5883 [hep-ph]} \BibitemShut
  {NoStop}%
\bibitem [{\citenamefont {Arkani-Hamed}\ \emph {et~al.}(2003)\citenamefont
  {Arkani-Hamed}, \citenamefont {Cheng}, \citenamefont {Creminelli},\ and\
  \citenamefont {Randall}}]{ArkaniHamed:2003wu}%
  \BibitemOpen
  \bibfield  {author} {\bibinfo {author} {\bibfnamefont {N.}~\bibnamefont
  {Arkani-Hamed}}, \bibinfo {author} {\bibfnamefont {H.-C.}\ \bibnamefont
  {Cheng}}, \bibinfo {author} {\bibfnamefont {P.}~\bibnamefont {Creminelli}}, \
  and\ \bibinfo {author} {\bibfnamefont {L.}~\bibnamefont {Randall}},\ }\href
  {\doibase 10.1103/PhysRevLett.90.221302} {\bibfield  {journal} {\bibinfo
  {journal} {Phys.Rev.Lett.}\ }\textbf {\bibinfo {volume} {90}},\ \bibinfo
  {pages} {221302} (\bibinfo {year} {2003})},\ \Eprint
  {http://arxiv.org/abs/hep-th/0301218} {arXiv:hep-th/0301218 [hep-th]}
  \BibitemShut {NoStop}%
\bibitem [{\citenamefont {Croon}\ and\ \citenamefont
  {Sanz}(2014)}]{Croon:2014dma}%
  \BibitemOpen
  \bibfield  {author} {\bibinfo {author} {\bibfnamefont {D.}~\bibnamefont
  {Croon}}\ and\ \bibinfo {author} {\bibfnamefont {V.}~\bibnamefont {Sanz}},\
  }\href@noop {} {\  (\bibinfo {year} {2014})},\ \Eprint
  {http://arxiv.org/abs/1411.7809} {arXiv:1411.7809 [hep-ph]} \BibitemShut
  {NoStop}%
\bibitem [{\citenamefont {Seiberg}\ and\ \citenamefont
  {Witten}(1994{\natexlab{a}})}]{Seiberg:1994rs}%
  \BibitemOpen
  \bibfield  {author} {\bibinfo {author} {\bibfnamefont {N.}~\bibnamefont
  {Seiberg}}\ and\ \bibinfo {author} {\bibfnamefont {E.}~\bibnamefont
  {Witten}},\ }\href {\doibase 10.1016/0550-3213(94)90124-4} {\bibfield
  {journal} {\bibinfo  {journal} {Nucl.Phys.}\ }\textbf {\bibinfo {volume}
  {B426}},\ \bibinfo {pages} {19} (\bibinfo {year} {1994}{\natexlab{a}})},\
  \Eprint {http://arxiv.org/abs/hep-th/9407087} {arXiv:hep-th/9407087 [hep-th]}
  \BibitemShut {NoStop}%
\bibitem [{\citenamefont {Seiberg}\ and\ \citenamefont
  {Witten}(1994{\natexlab{b}})}]{Seiberg:1994aj}%
  \BibitemOpen
  \bibfield  {author} {\bibinfo {author} {\bibfnamefont {N.}~\bibnamefont
  {Seiberg}}\ and\ \bibinfo {author} {\bibfnamefont {E.}~\bibnamefont
  {Witten}},\ }\href {\doibase 10.1016/0550-3213(94)90214-3} {\bibfield
  {journal} {\bibinfo  {journal} {Nucl.Phys.}\ }\textbf {\bibinfo {volume}
  {B431}},\ \bibinfo {pages} {484} (\bibinfo {year} {1994}{\natexlab{b}})},\
  \Eprint {http://arxiv.org/abs/hep-th/9408099} {arXiv:hep-th/9408099 [hep-th]}
  \BibitemShut {NoStop}%
\bibitem [{\citenamefont {Berg}\ \emph
  {et~al.}(2005{\natexlab{a}})\citenamefont {Berg}, \citenamefont {Haack},\
  and\ \citenamefont {Kors}}]{Berg:2004ek}%
  \BibitemOpen
  \bibfield  {author} {\bibinfo {author} {\bibfnamefont {M.}~\bibnamefont
  {Berg}}, \bibinfo {author} {\bibfnamefont {M.}~\bibnamefont {Haack}}, \ and\
  \bibinfo {author} {\bibfnamefont {B.}~\bibnamefont {Kors}},\ }\href {\doibase
  10.1103/PhysRevD.71.026005} {\bibfield  {journal} {\bibinfo  {journal}
  {Phys.Rev.}\ }\textbf {\bibinfo {volume} {D71}},\ \bibinfo {pages} {026005}
  (\bibinfo {year} {2005}{\natexlab{a}})},\ \Eprint
  {http://arxiv.org/abs/hep-th/0404087} {arXiv:hep-th/0404087 [hep-th]}
  \BibitemShut {NoStop}%
\bibitem [{\citenamefont {Baumann}\ \emph {et~al.}(2006)\citenamefont
  {Baumann}, \citenamefont {Dymarsky}, \citenamefont {Klebanov}, \citenamefont
  {Maldacena}, \citenamefont {McAllister} \emph {et~al.}}]{Baumann:2006th}%
  \BibitemOpen
  \bibfield  {author} {\bibinfo {author} {\bibfnamefont {D.}~\bibnamefont
  {Baumann}}, \bibinfo {author} {\bibfnamefont {A.}~\bibnamefont {Dymarsky}},
  \bibinfo {author} {\bibfnamefont {I.~R.}\ \bibnamefont {Klebanov}}, \bibinfo
  {author} {\bibfnamefont {J.~M.}\ \bibnamefont {Maldacena}}, \bibinfo {author}
  {\bibfnamefont {L.~P.}\ \bibnamefont {McAllister}},  \emph {et~al.},\ }\href
  {\doibase 10.1088/1126-6708/2006/11/031} {\bibfield  {journal} {\bibinfo
  {journal} {JHEP}\ }\textbf {\bibinfo {volume} {0611}},\ \bibinfo {pages}
  {031} (\bibinfo {year} {2006})},\ \Eprint
  {http://arxiv.org/abs/hep-th/0607050} {arXiv:hep-th/0607050 [hep-th]}
  \BibitemShut {NoStop}%
\bibitem [{\citenamefont {Cremades}\ \emph {et~al.}(2003)\citenamefont
  {Cremades}, \citenamefont {Ibanez},\ and\ \citenamefont
  {Marchesano}}]{Cremades:2003qj}%
  \BibitemOpen
  \bibfield  {author} {\bibinfo {author} {\bibfnamefont {D.}~\bibnamefont
  {Cremades}}, \bibinfo {author} {\bibfnamefont {L.}~\bibnamefont {Ibanez}}, \
  and\ \bibinfo {author} {\bibfnamefont {F.}~\bibnamefont {Marchesano}},\
  }\href {\doibase 10.1088/1126-6708/2003/07/038} {\bibfield  {journal}
  {\bibinfo  {journal} {JHEP}\ }\textbf {\bibinfo {volume} {0307}},\ \bibinfo
  {pages} {038} (\bibinfo {year} {2003})},\ \Eprint
  {http://arxiv.org/abs/hep-th/0302105} {arXiv:hep-th/0302105 [hep-th]}
  \BibitemShut {NoStop}%
\bibitem [{\citenamefont {Cremades}\ \emph {et~al.}(2004)\citenamefont
  {Cremades}, \citenamefont {Ibanez},\ and\ \citenamefont
  {Marchesano}}]{Cremades:2004wa}%
  \BibitemOpen
  \bibfield  {author} {\bibinfo {author} {\bibfnamefont {D.}~\bibnamefont
  {Cremades}}, \bibinfo {author} {\bibfnamefont {L.}~\bibnamefont {Ibanez}}, \
  and\ \bibinfo {author} {\bibfnamefont {F.}~\bibnamefont {Marchesano}},\
  }\href {\doibase 10.1088/1126-6708/2004/05/079} {\bibfield  {journal}
  {\bibinfo  {journal} {JHEP}\ }\textbf {\bibinfo {volume} {0405}},\ \bibinfo
  {pages} {079} (\bibinfo {year} {2004})},\ \Eprint
  {http://arxiv.org/abs/hep-th/0404229} {arXiv:hep-th/0404229 [hep-th]}
  \BibitemShut {NoStop}%
\bibitem [{\citenamefont {Marino}\ \emph {et~al.}(2000)\citenamefont {Marino},
  \citenamefont {Minasian}, \citenamefont {Moore},\ and\ \citenamefont
  {Strominger}}]{Marino:1999af}%
  \BibitemOpen
  \bibfield  {author} {\bibinfo {author} {\bibfnamefont {M.}~\bibnamefont
  {Marino}}, \bibinfo {author} {\bibfnamefont {R.}~\bibnamefont {Minasian}},
  \bibinfo {author} {\bibfnamefont {G.~W.}\ \bibnamefont {Moore}}, \ and\
  \bibinfo {author} {\bibfnamefont {A.}~\bibnamefont {Strominger}},\ }\href
  {\doibase 10.1088/1126-6708/2000/01/005} {\bibfield  {journal} {\bibinfo
  {journal} {JHEP}\ }\textbf {\bibinfo {volume} {0001}},\ \bibinfo {pages}
  {005} (\bibinfo {year} {2000})},\ \Eprint
  {http://arxiv.org/abs/hep-th/9911206} {arXiv:hep-th/9911206 [hep-th]}
  \BibitemShut {NoStop}%
\bibitem [{\citenamefont {Hamidi}\ and\ \citenamefont
  {Vafa}(1987)}]{Hamidi:1986vh}%
  \BibitemOpen
  \bibfield  {author} {\bibinfo {author} {\bibfnamefont {S.}~\bibnamefont
  {Hamidi}}\ and\ \bibinfo {author} {\bibfnamefont {C.}~\bibnamefont {Vafa}},\
  }\href {\doibase 10.1016/0550-3213(87)90006-X} {\bibfield  {journal}
  {\bibinfo  {journal} {Nucl.Phys.}\ }\textbf {\bibinfo {volume} {B279}},\
  \bibinfo {pages} {465} (\bibinfo {year} {1987})}\BibitemShut {NoStop}%
\bibitem [{\citenamefont {Dundee}\ \emph {et~al.}(2010)\citenamefont {Dundee},
  \citenamefont {Raby},\ and\ \citenamefont {Westphal}}]{Dundee:2010sb}%
  \BibitemOpen
  \bibfield  {author} {\bibinfo {author} {\bibfnamefont {B.}~\bibnamefont
  {Dundee}}, \bibinfo {author} {\bibfnamefont {S.}~\bibnamefont {Raby}}, \ and\
  \bibinfo {author} {\bibfnamefont {A.}~\bibnamefont {Westphal}},\ }\href
  {\doibase 10.1103/PhysRevD.82.126002} {\bibfield  {journal} {\bibinfo
  {journal} {Phys.Rev.}\ }\textbf {\bibinfo {volume} {D82}},\ \bibinfo {pages}
  {126002} (\bibinfo {year} {2010})},\ \Eprint {http://arxiv.org/abs/1002.1081}
  {arXiv:1002.1081 [hep-th]} \BibitemShut {NoStop}%
\bibitem [{\citenamefont {Abe}\ \emph {et~al.}(2014{\natexlab{a}})\citenamefont
  {Abe}, \citenamefont {Kobayashi},\ and\ \citenamefont
  {Otsuka}}]{Abe:2014xja}%
  \BibitemOpen
  \bibfield  {author} {\bibinfo {author} {\bibfnamefont {H.}~\bibnamefont
  {Abe}}, \bibinfo {author} {\bibfnamefont {T.}~\bibnamefont {Kobayashi}}, \
  and\ \bibinfo {author} {\bibfnamefont {H.}~\bibnamefont {Otsuka}},\
  }\href@noop {} {\  (\bibinfo {year} {2014}{\natexlab{a}})},\ \Eprint
  {http://arxiv.org/abs/1411.4768} {arXiv:1411.4768 [hep-th]} \BibitemShut
  {NoStop}%
\bibitem [{\citenamefont {Schimmrigk}(2014)}]{Schimmrigk:2014ica}%
  \BibitemOpen
  \bibfield  {author} {\bibinfo {author} {\bibfnamefont {R.}~\bibnamefont
  {Schimmrigk}},\ }\href@noop {} {\  (\bibinfo {year} {2014})},\ \Eprint
  {http://arxiv.org/abs/1412.8537} {arXiv:1412.8537 [hep-th]} \BibitemShut
  {NoStop}%
\bibitem [{\citenamefont {Kobayashi}\ and\ \citenamefont
  {Takahashi}(2011)}]{Kobayashi:2010pz}%
  \BibitemOpen
  \bibfield  {author} {\bibinfo {author} {\bibfnamefont {T.}~\bibnamefont
  {Kobayashi}}\ and\ \bibinfo {author} {\bibfnamefont {F.}~\bibnamefont
  {Takahashi}},\ }\href {\doibase 10.1088/1475-7516/2011/01/026} {\bibfield
  {journal} {\bibinfo  {journal} {JCAP}\ }\textbf {\bibinfo {volume} {1101}},\
  \bibinfo {pages} {026} (\bibinfo {year} {2011})},\ \Eprint
  {http://arxiv.org/abs/1011.3988} {arXiv:1011.3988 [astro-ph.CO]} \BibitemShut
  {NoStop}%
\bibitem [{\citenamefont {Feng}\ \emph {et~al.}(2003)\citenamefont {Feng},
  \citenamefont {Li}, \citenamefont {Zhang},\ and\ \citenamefont
  {Zhang}}]{Feng:2003mk}%
  \BibitemOpen
  \bibfield  {author} {\bibinfo {author} {\bibfnamefont {B.}~\bibnamefont
  {Feng}}, \bibinfo {author} {\bibfnamefont {M.-z.}\ \bibnamefont {Li}},
  \bibinfo {author} {\bibfnamefont {R.-J.}\ \bibnamefont {Zhang}}, \ and\
  \bibinfo {author} {\bibfnamefont {X.-m.}\ \bibnamefont {Zhang}},\ }\href
  {\doibase 10.1103/PhysRevD.68.103511} {\bibfield  {journal} {\bibinfo
  {journal} {Phys.Rev.}\ }\textbf {\bibinfo {volume} {D68}},\ \bibinfo {pages}
  {103511} (\bibinfo {year} {2003})},\ \Eprint
  {http://arxiv.org/abs/astro-ph/0302479} {arXiv:astro-ph/0302479 [astro-ph]}
  \BibitemShut {NoStop}%
\bibitem [{\citenamefont {Takahashi}(2013)}]{Takahashi:2013tj}%
  \BibitemOpen
  \bibfield  {author} {\bibinfo {author} {\bibfnamefont {F.}~\bibnamefont
  {Takahashi}},\ }\href {\doibase 10.1088/1475-7516/2013/06/013} {\bibfield
  {journal} {\bibinfo  {journal} {JCAP}\ }\textbf {\bibinfo {volume} {1306}},\
  \bibinfo {pages} {013} (\bibinfo {year} {2013})},\ \Eprint
  {http://arxiv.org/abs/1301.2834} {arXiv:1301.2834} \BibitemShut {NoStop}%
\bibitem [{\citenamefont {Czerny}\ \emph
  {et~al.}(2014{\natexlab{c}})\citenamefont {Czerny}, \citenamefont
  {Kobayashi},\ and\ \citenamefont {Takahashi}}]{Czerny:2014wua}%
  \BibitemOpen
  \bibfield  {author} {\bibinfo {author} {\bibfnamefont {M.}~\bibnamefont
  {Czerny}}, \bibinfo {author} {\bibfnamefont {T.}~\bibnamefont {Kobayashi}}, \
  and\ \bibinfo {author} {\bibfnamefont {F.}~\bibnamefont {Takahashi}},\ }\href
  {\doibase 10.1016/j.physletb.2014.06.018} {\  (\bibinfo {year}
  {2014}{\natexlab{c}}),\ 10.1016/j.physletb.2014.06.018},\ \Eprint
  {http://arxiv.org/abs/1403.4589} {arXiv:1403.4589 [astro-ph.CO]} \BibitemShut
  {NoStop}%
\bibitem [{\citenamefont {Abazajian}\ \emph {et~al.}(2014)\citenamefont
  {Abazajian}, \citenamefont {Aslanyan}, \citenamefont {Easther},\ and\
  \citenamefont {Price}}]{Abazajian:2014tqa}%
  \BibitemOpen
  \bibfield  {author} {\bibinfo {author} {\bibfnamefont {K.~N.}\ \bibnamefont
  {Abazajian}}, \bibinfo {author} {\bibfnamefont {G.}~\bibnamefont {Aslanyan}},
  \bibinfo {author} {\bibfnamefont {R.}~\bibnamefont {Easther}}, \ and\
  \bibinfo {author} {\bibfnamefont {L.~C.}\ \bibnamefont {Price}},\ }\href
  {\doibase 10.1088/1475-7516/2014/08/053} {\bibfield  {journal} {\bibinfo
  {journal} {JCAP}\ }\textbf {\bibinfo {volume} {1408}},\ \bibinfo {pages}
  {053} (\bibinfo {year} {2014})},\ \Eprint {http://arxiv.org/abs/1403.5922}
  {arXiv:1403.5922 [astro-ph.CO]} \BibitemShut {NoStop}%
\bibitem [{\citenamefont {Wan}\ \emph {et~al.}(2014)\citenamefont {Wan},
  \citenamefont {Li}, \citenamefont {Li}, \citenamefont {Qiu}, \citenamefont
  {Cai} \emph {et~al.}}]{Wan:2014fra}%
  \BibitemOpen
  \bibfield  {author} {\bibinfo {author} {\bibfnamefont {Y.}~\bibnamefont
  {Wan}}, \bibinfo {author} {\bibfnamefont {S.}~\bibnamefont {Li}}, \bibinfo
  {author} {\bibfnamefont {M.}~\bibnamefont {Li}}, \bibinfo {author}
  {\bibfnamefont {T.}~\bibnamefont {Qiu}}, \bibinfo {author} {\bibfnamefont
  {Y.}~\bibnamefont {Cai}},  \emph {et~al.},\ }\href {\doibase
  10.1103/PhysRevD.90.023537} {\bibfield  {journal} {\bibinfo  {journal}
  {Phys.Rev.}\ }\textbf {\bibinfo {volume} {D90}},\ \bibinfo {pages} {023537}
  (\bibinfo {year} {2014})},\ \Eprint {http://arxiv.org/abs/1405.2784}
  {arXiv:1405.2784 [astro-ph.CO]} \BibitemShut {NoStop}%
\bibitem [{\citenamefont {Minor}\ and\ \citenamefont
  {Kaplinghat}(2014)}]{Minor:2014xla}%
  \BibitemOpen
  \bibfield  {author} {\bibinfo {author} {\bibfnamefont {Q.~E.}\ \bibnamefont
  {Minor}}\ and\ \bibinfo {author} {\bibfnamefont {M.}~\bibnamefont
  {Kaplinghat}},\ }\href@noop {} {\  (\bibinfo {year} {2014})},\ \Eprint
  {http://arxiv.org/abs/1411.0689} {arXiv:1411.0689 [astro-ph.CO]} \BibitemShut
  {NoStop}%
\bibitem [{\citenamefont {de~la Fuente}\ \emph {et~al.}(2014)\citenamefont
  {de~la Fuente}, \citenamefont {Saraswat},\ and\ \citenamefont
  {Sundrum}}]{delaFuente:2014aca}%
  \BibitemOpen
  \bibfield  {author} {\bibinfo {author} {\bibfnamefont {A.}~\bibnamefont
  {de~la Fuente}}, \bibinfo {author} {\bibfnamefont {P.}~\bibnamefont
  {Saraswat}}, \ and\ \bibinfo {author} {\bibfnamefont {R.}~\bibnamefont
  {Sundrum}},\ }\href@noop {} {\  (\bibinfo {year} {2014})},\ \Eprint
  {http://arxiv.org/abs/1412.3457} {arXiv:1412.3457 [hep-th]} \BibitemShut
  {NoStop}%
\bibitem [{\citenamefont {Ade}\ \emph {et~al.}(2014{\natexlab{b}})\citenamefont
  {Ade} \emph {et~al.}}]{Ade:2013zuv}%
  \BibitemOpen
  \bibfield  {author} {\bibinfo {author} {\bibfnamefont {P.}~\bibnamefont
  {Ade}} \emph {et~al.} (\bibinfo {collaboration} {Planck Collaboration}),\
  }\href {\doibase 10.1051/0004-6361/201321591} {\bibfield  {journal} {\bibinfo
   {journal} {Astron.Astrophys.}\ }\textbf {\bibinfo {volume} {571}},\ \bibinfo
  {pages} {A16} (\bibinfo {year} {2014}{\natexlab{b}})},\ \Eprint
  {http://arxiv.org/abs/1303.5076} {arXiv:1303.5076 [astro-ph.CO]} \BibitemShut
  {NoStop}%
\bibitem [{\citenamefont {Easther}\ and\ \citenamefont
  {Peiris}(2006)}]{Easther:2006tv}%
  \BibitemOpen
  \bibfield  {author} {\bibinfo {author} {\bibfnamefont {R.}~\bibnamefont
  {Easther}}\ and\ \bibinfo {author} {\bibfnamefont {H.}~\bibnamefont
  {Peiris}},\ }\href {\doibase 10.1088/1475-7516/2006/09/010} {\bibfield
  {journal} {\bibinfo  {journal} {JCAP}\ }\textbf {\bibinfo {volume} {0609}},\
  \bibinfo {pages} {010} (\bibinfo {year} {2006})},\ \Eprint
  {http://arxiv.org/abs/astro-ph/0604214} {arXiv:astro-ph/0604214 [astro-ph]}
  \BibitemShut {NoStop}%
\bibitem [{\citenamefont {Abe}\ \emph {et~al.}(2014{\natexlab{b}})\citenamefont
  {Abe}, \citenamefont {Kobayashi}, \citenamefont {Sumita},\ and\ \citenamefont
  {Tatsuta}}]{Abe:2014vza}%
  \BibitemOpen
  \bibfield  {author} {\bibinfo {author} {\bibfnamefont {H.}~\bibnamefont
  {Abe}}, \bibinfo {author} {\bibfnamefont {T.}~\bibnamefont {Kobayashi}},
  \bibinfo {author} {\bibfnamefont {K.}~\bibnamefont {Sumita}}, \ and\ \bibinfo
  {author} {\bibfnamefont {Y.}~\bibnamefont {Tatsuta}},\ }\href {\doibase
  10.1103/PhysRevD.90.105006} {\bibfield  {journal} {\bibinfo  {journal}
  {Phys.Rev.}\ }\textbf {\bibinfo {volume} {D90}},\ \bibinfo {pages} {105006}
  (\bibinfo {year} {2014}{\natexlab{b}})},\ \Eprint
  {http://arxiv.org/abs/1405.5012} {arXiv:1405.5012 [hep-ph]} \BibitemShut
  {NoStop}%
\bibitem [{\citenamefont {Polchinski}(1998)}]{Polchinski:1998rq}%
  \BibitemOpen
  \bibfield  {author} {\bibinfo {author} {\bibfnamefont {J.}~\bibnamefont
  {Polchinski}},\ }\href@noop {} {\  (\bibinfo {year} {1998})}\BibitemShut
  {NoStop}%
\bibitem [{\citenamefont {Antoniadis}\ \emph {et~al.}(2001)\citenamefont
  {Antoniadis}, \citenamefont {Benakli},\ and\ \citenamefont
  {Quiros}}]{Antoniadis:2001cv}%
  \BibitemOpen
  \bibfield  {author} {\bibinfo {author} {\bibfnamefont {I.}~\bibnamefont
  {Antoniadis}}, \bibinfo {author} {\bibfnamefont {K.}~\bibnamefont {Benakli}},
  \ and\ \bibinfo {author} {\bibfnamefont {M.}~\bibnamefont {Quiros}},\ }\href
  {\doibase 10.1088/1367-2630/3/1/320} {\bibfield  {journal} {\bibinfo
  {journal} {New J.Phys.}\ }\textbf {\bibinfo {volume} {3}},\ \bibinfo {pages}
  {20} (\bibinfo {year} {2001})},\ \Eprint
  {http://arxiv.org/abs/hep-th/0108005} {arXiv:hep-th/0108005 [hep-th]}
  \BibitemShut {NoStop}%
\bibitem [{\citenamefont {Blumenhagen}\ \emph {et~al.}(2007)\citenamefont
  {Blumenhagen}, \citenamefont {Kors}, \citenamefont {Lust},\ and\
  \citenamefont {Stieberger}}]{Blumenhagen:2006ci}%
  \BibitemOpen
  \bibfield  {author} {\bibinfo {author} {\bibfnamefont {R.}~\bibnamefont
  {Blumenhagen}}, \bibinfo {author} {\bibfnamefont {B.}~\bibnamefont {Kors}},
  \bibinfo {author} {\bibfnamefont {D.}~\bibnamefont {Lust}}, \ and\ \bibinfo
  {author} {\bibfnamefont {S.}~\bibnamefont {Stieberger}},\ }\href {\doibase
  10.1016/j.physrep.2007.04.003} {\bibfield  {journal} {\bibinfo  {journal}
  {Phys.Rept.}\ }\textbf {\bibinfo {volume} {445}},\ \bibinfo {pages} {1}
  (\bibinfo {year} {2007})},\ \Eprint {http://arxiv.org/abs/hep-th/0610327}
  {arXiv:hep-th/0610327 [hep-th]} \BibitemShut {NoStop}%
\bibitem [{\citenamefont {Jockers}\ and\ \citenamefont
  {Louis}(2005)}]{Jockers:2004yj}%
  \BibitemOpen
  \bibfield  {author} {\bibinfo {author} {\bibfnamefont {H.}~\bibnamefont
  {Jockers}}\ and\ \bibinfo {author} {\bibfnamefont {J.}~\bibnamefont
  {Louis}},\ }\href {\doibase 10.1016/j.nuclphysb.2004.11.009} {\bibfield
  {journal} {\bibinfo  {journal} {Nucl.Phys.}\ }\textbf {\bibinfo {volume}
  {B705}},\ \bibinfo {pages} {167} (\bibinfo {year} {2005})},\ \Eprint
  {http://arxiv.org/abs/hep-th/0409098} {arXiv:hep-th/0409098 [hep-th]}
  \BibitemShut {NoStop}%
\bibitem [{\citenamefont {Izawa}\ and\ \citenamefont
  {Yanagida}(1996)}]{Izawa:1996pk}%
  \BibitemOpen
  \bibfield  {author} {\bibinfo {author} {\bibfnamefont {K.-I.}\ \bibnamefont
  {Izawa}}\ and\ \bibinfo {author} {\bibfnamefont {T.}~\bibnamefont
  {Yanagida}},\ }\href {\doibase 10.1143/PTP.95.829} {\bibfield  {journal}
  {\bibinfo  {journal} {Prog.Theor.Phys.}\ }\textbf {\bibinfo {volume} {95}},\
  \bibinfo {pages} {829} (\bibinfo {year} {1996})},\ \Eprint
  {http://arxiv.org/abs/hep-th/9602180} {arXiv:hep-th/9602180 [hep-th]}
  \BibitemShut {NoStop}%
\bibitem [{\citenamefont {Intriligator}\ and\ \citenamefont
  {Thomas}(1996)}]{Intriligator:1996pu}%
  \BibitemOpen
  \bibfield  {author} {\bibinfo {author} {\bibfnamefont {K.~A.}\ \bibnamefont
  {Intriligator}}\ and\ \bibinfo {author} {\bibfnamefont {S.~D.}\ \bibnamefont
  {Thomas}},\ }\href {\doibase 10.1016/0550-3213(96)00261-1} {\bibfield
  {journal} {\bibinfo  {journal} {Nucl.Phys.}\ }\textbf {\bibinfo {volume}
  {B473}},\ \bibinfo {pages} {121} (\bibinfo {year} {1996})},\ \Eprint
  {http://arxiv.org/abs/hep-th/9603158} {arXiv:hep-th/9603158 [hep-th]}
  \BibitemShut {NoStop}%
\bibitem [{\citenamefont {Kachru}\ \emph
  {et~al.}(2003{\natexlab{a}})\citenamefont {Kachru}, \citenamefont {Kallosh},
  \citenamefont {Linde},\ and\ \citenamefont {Trivedi}}]{Kachru:2003aw}%
  \BibitemOpen
  \bibfield  {author} {\bibinfo {author} {\bibfnamefont {S.}~\bibnamefont
  {Kachru}}, \bibinfo {author} {\bibfnamefont {R.}~\bibnamefont {Kallosh}},
  \bibinfo {author} {\bibfnamefont {A.~D.}\ \bibnamefont {Linde}}, \ and\
  \bibinfo {author} {\bibfnamefont {S.~P.}\ \bibnamefont {Trivedi}},\ }\href
  {\doibase 10.1103/PhysRevD.68.046005} {\bibfield  {journal} {\bibinfo
  {journal} {Phys.Rev.}\ }\textbf {\bibinfo {volume} {D68}},\ \bibinfo {pages}
  {046005} (\bibinfo {year} {2003}{\natexlab{a}})},\ \Eprint
  {http://arxiv.org/abs/hep-th/0301240} {arXiv:hep-th/0301240 [hep-th]}
  \BibitemShut {NoStop}%
\bibitem [{\citenamefont {Grimm}\ and\ \citenamefont
  {Louis}(2004)}]{Grimm:2004uq}%
  \BibitemOpen
  \bibfield  {author} {\bibinfo {author} {\bibfnamefont {T.~W.}\ \bibnamefont
  {Grimm}}\ and\ \bibinfo {author} {\bibfnamefont {J.}~\bibnamefont {Louis}},\
  }\href {\doibase 10.1016/j.nuclphysb.2004.08.005} {\bibfield  {journal}
  {\bibinfo  {journal} {Nucl.Phys.}\ }\textbf {\bibinfo {volume} {B699}},\
  \bibinfo {pages} {387} (\bibinfo {year} {2004})},\ \Eprint
  {http://arxiv.org/abs/hep-th/0403067} {arXiv:hep-th/0403067 [hep-th]}
  \BibitemShut {NoStop}%
\bibitem [{\citenamefont {Balasubramanian}\ \emph {et~al.}(2005)\citenamefont
  {Balasubramanian}, \citenamefont {Berglund}, \citenamefont {Conlon},\ and\
  \citenamefont {Quevedo}}]{Balasubramanian:2005zx}%
  \BibitemOpen
  \bibfield  {author} {\bibinfo {author} {\bibfnamefont {V.}~\bibnamefont
  {Balasubramanian}}, \bibinfo {author} {\bibfnamefont {P.}~\bibnamefont
  {Berglund}}, \bibinfo {author} {\bibfnamefont {J.~P.}\ \bibnamefont
  {Conlon}}, \ and\ \bibinfo {author} {\bibfnamefont {F.}~\bibnamefont
  {Quevedo}},\ }\href {\doibase 10.1088/1126-6708/2005/03/007} {\bibfield
  {journal} {\bibinfo  {journal} {JHEP}\ }\textbf {\bibinfo {volume} {0503}},\
  \bibinfo {pages} {007} (\bibinfo {year} {2005})},\ \Eprint
  {http://arxiv.org/abs/hep-th/0502058} {arXiv:hep-th/0502058 [hep-th]}
  \BibitemShut {NoStop}%
\bibitem [{\citenamefont {Conlon}\ \emph {et~al.}(2005)\citenamefont {Conlon},
  \citenamefont {Quevedo},\ and\ \citenamefont {Suruliz}}]{Conlon:2005ki}%
  \BibitemOpen
  \bibfield  {author} {\bibinfo {author} {\bibfnamefont {J.~P.}\ \bibnamefont
  {Conlon}}, \bibinfo {author} {\bibfnamefont {F.}~\bibnamefont {Quevedo}}, \
  and\ \bibinfo {author} {\bibfnamefont {K.}~\bibnamefont {Suruliz}},\ }\href
  {\doibase 10.1088/1126-6708/2005/08/007} {\bibfield  {journal} {\bibinfo
  {journal} {JHEP}\ }\textbf {\bibinfo {volume} {0508}},\ \bibinfo {pages}
  {007} (\bibinfo {year} {2005})},\ \Eprint
  {http://arxiv.org/abs/hep-th/0505076} {arXiv:hep-th/0505076 [hep-th]}
  \BibitemShut {NoStop}%
\bibitem [{\citenamefont {Choi}\ and\ \citenamefont
  {Jeong}(2006)}]{Choi:2006bh}%
  \BibitemOpen
  \bibfield  {author} {\bibinfo {author} {\bibfnamefont {K.}~\bibnamefont
  {Choi}}\ and\ \bibinfo {author} {\bibfnamefont {K.~S.}\ \bibnamefont
  {Jeong}},\ }\href {\doibase 10.1088/1126-6708/2006/08/007} {\bibfield
  {journal} {\bibinfo  {journal} {JHEP}\ }\textbf {\bibinfo {volume} {0608}},\
  \bibinfo {pages} {007} (\bibinfo {year} {2006})},\ \Eprint
  {http://arxiv.org/abs/hep-th/0605108} {arXiv:hep-th/0605108 [hep-th]}
  \BibitemShut {NoStop}%
\bibitem [{\citenamefont {Haack}\ \emph {et~al.}(2007)\citenamefont {Haack},
  \citenamefont {Krefl}, \citenamefont {Lust}, \citenamefont {Van~Proeyen},\
  and\ \citenamefont {Zagermann}}]{Haack:2006cy}%
  \BibitemOpen
  \bibfield  {author} {\bibinfo {author} {\bibfnamefont {M.}~\bibnamefont
  {Haack}}, \bibinfo {author} {\bibfnamefont {D.}~\bibnamefont {Krefl}},
  \bibinfo {author} {\bibfnamefont {D.}~\bibnamefont {Lust}}, \bibinfo {author}
  {\bibfnamefont {A.}~\bibnamefont {Van~Proeyen}}, \ and\ \bibinfo {author}
  {\bibfnamefont {M.}~\bibnamefont {Zagermann}},\ }\href {\doibase
  10.1088/1126-6708/2007/01/078} {\bibfield  {journal} {\bibinfo  {journal}
  {JHEP}\ }\textbf {\bibinfo {volume} {0701}},\ \bibinfo {pages} {078}
  (\bibinfo {year} {2007})},\ \Eprint {http://arxiv.org/abs/hep-th/0609211}
  {arXiv:hep-th/0609211 [hep-th]} \BibitemShut {NoStop}%
\bibitem [{\citenamefont {Camara}\ \emph {et~al.}(2004)\citenamefont {Camara},
  \citenamefont {Ibanez},\ and\ \citenamefont {Uranga}}]{Camara:2003ku}%
  \BibitemOpen
  \bibfield  {author} {\bibinfo {author} {\bibfnamefont {P.~G.}\ \bibnamefont
  {Camara}}, \bibinfo {author} {\bibfnamefont {L.}~\bibnamefont {Ibanez}}, \
  and\ \bibinfo {author} {\bibfnamefont {A.}~\bibnamefont {Uranga}},\ }\href
  {\doibase 10.1016/j.nuclphysb.2004.04.013} {\bibfield  {journal} {\bibinfo
  {journal} {Nucl.Phys.}\ }\textbf {\bibinfo {volume} {B689}},\ \bibinfo
  {pages} {195} (\bibinfo {year} {2004})},\ \Eprint
  {http://arxiv.org/abs/hep-th/0311241} {arXiv:hep-th/0311241 [hep-th]}
  \BibitemShut {NoStop}%
\bibitem [{\citenamefont {Camara}\ \emph {et~al.}(2005)\citenamefont {Camara},
  \citenamefont {Ibanez},\ and\ \citenamefont {Uranga}}]{Camara:2004jj}%
  \BibitemOpen
  \bibfield  {author} {\bibinfo {author} {\bibfnamefont {P.~G.}\ \bibnamefont
  {Camara}}, \bibinfo {author} {\bibfnamefont {L.}~\bibnamefont {Ibanez}}, \
  and\ \bibinfo {author} {\bibfnamefont {A.}~\bibnamefont {Uranga}},\ }\href
  {\doibase 10.1016/j.nuclphysb.2004.11.035} {\bibfield  {journal} {\bibinfo
  {journal} {Nucl.Phys.}\ }\textbf {\bibinfo {volume} {B708}},\ \bibinfo
  {pages} {268} (\bibinfo {year} {2005})},\ \Eprint
  {http://arxiv.org/abs/hep-th/0408036} {arXiv:hep-th/0408036 [hep-th]}
  \BibitemShut {NoStop}%
\bibitem [{\citenamefont {Berg}\ \emph
  {et~al.}(2005{\natexlab{b}})\citenamefont {Berg}, \citenamefont {Haack},\
  and\ \citenamefont {Kors}}]{Berg:2005ja}%
  \BibitemOpen
  \bibfield  {author} {\bibinfo {author} {\bibfnamefont {M.}~\bibnamefont
  {Berg}}, \bibinfo {author} {\bibfnamefont {M.}~\bibnamefont {Haack}}, \ and\
  \bibinfo {author} {\bibfnamefont {B.}~\bibnamefont {Kors}},\ }\href {\doibase
  10.1088/1126-6708/2005/11/030} {\bibfield  {journal} {\bibinfo  {journal}
  {JHEP}\ }\textbf {\bibinfo {volume} {0511}},\ \bibinfo {pages} {030}
  (\bibinfo {year} {2005}{\natexlab{b}})},\ \Eprint
  {http://arxiv.org/abs/hep-th/0508043} {arXiv:hep-th/0508043 [hep-th]}
  \BibitemShut {NoStop}%
\bibitem [{\citenamefont {Collinucci}\ \emph {et~al.}(2009)\citenamefont
  {Collinucci}, \citenamefont {Denef},\ and\ \citenamefont
  {Esole}}]{Collinucci:2008pf}%
  \BibitemOpen
  \bibfield  {author} {\bibinfo {author} {\bibfnamefont {A.}~\bibnamefont
  {Collinucci}}, \bibinfo {author} {\bibfnamefont {F.}~\bibnamefont {Denef}}, \
  and\ \bibinfo {author} {\bibfnamefont {M.}~\bibnamefont {Esole}},\ }\href
  {\doibase 10.1088/1126-6708/2009/02/005} {\bibfield  {journal} {\bibinfo
  {journal} {JHEP}\ }\textbf {\bibinfo {volume} {0902}},\ \bibinfo {pages}
  {005} (\bibinfo {year} {2009})},\ \Eprint {http://arxiv.org/abs/0805.1573}
  {arXiv:0805.1573 [hep-th]} \BibitemShut {NoStop}%
\bibitem [{\citenamefont {Abe}\ \emph {et~al.}(2006)\citenamefont {Abe},
  \citenamefont {Higaki},\ and\ \citenamefont {Kobayashi}}]{Abe:2006xi}%
  \BibitemOpen
  \bibfield  {author} {\bibinfo {author} {\bibfnamefont {H.}~\bibnamefont
  {Abe}}, \bibinfo {author} {\bibfnamefont {T.}~\bibnamefont {Higaki}}, \ and\
  \bibinfo {author} {\bibfnamefont {T.}~\bibnamefont {Kobayashi}},\ }\href
  {\doibase 10.1103/PhysRevD.74.045012} {\bibfield  {journal} {\bibinfo
  {journal} {Phys.Rev.}\ }\textbf {\bibinfo {volume} {D74}},\ \bibinfo {pages}
  {045012} (\bibinfo {year} {2006})},\ \Eprint
  {http://arxiv.org/abs/hep-th/0606095} {arXiv:hep-th/0606095 [hep-th]}
  \BibitemShut {NoStop}%
\bibitem [{\citenamefont {DeWolfe}\ \emph {et~al.}(2005)\citenamefont
  {DeWolfe}, \citenamefont {Giryavets}, \citenamefont {Kachru},\ and\
  \citenamefont {Taylor}}]{DeWolfe:2005uu}%
  \BibitemOpen
  \bibfield  {author} {\bibinfo {author} {\bibfnamefont {O.}~\bibnamefont
  {DeWolfe}}, \bibinfo {author} {\bibfnamefont {A.}~\bibnamefont {Giryavets}},
  \bibinfo {author} {\bibfnamefont {S.}~\bibnamefont {Kachru}}, \ and\ \bibinfo
  {author} {\bibfnamefont {W.}~\bibnamefont {Taylor}},\ }\href {\doibase
  10.1088/1126-6708/2005/07/066} {\bibfield  {journal} {\bibinfo  {journal}
  {JHEP}\ }\textbf {\bibinfo {volume} {0507}},\ \bibinfo {pages} {066}
  (\bibinfo {year} {2005})},\ \Eprint {http://arxiv.org/abs/hep-th/0505160}
  {arXiv:hep-th/0505160 [hep-th]} \BibitemShut {NoStop}%
\bibitem [{\citenamefont {Grimm}(2007)}]{Grimm:2007xm}%
  \BibitemOpen
  \bibfield  {author} {\bibinfo {author} {\bibfnamefont {T.~W.}\ \bibnamefont
  {Grimm}},\ }\href {\doibase 10.1088/1126-6708/2007/10/004} {\bibfield
  {journal} {\bibinfo  {journal} {JHEP}\ }\textbf {\bibinfo {volume} {0710}},\
  \bibinfo {pages} {004} (\bibinfo {year} {2007})},\ \Eprint
  {http://arxiv.org/abs/0705.3253} {arXiv:0705.3253 [hep-th]} \BibitemShut
  {NoStop}%
\bibitem [{\citenamefont {Grimm}(2008)}]{Grimm:2007hs}%
  \BibitemOpen
  \bibfield  {author} {\bibinfo {author} {\bibfnamefont {T.~W.}\ \bibnamefont
  {Grimm}},\ }\href {\doibase 10.1103/PhysRevD.77.126007} {\bibfield  {journal}
  {\bibinfo  {journal} {Phys.Rev.}\ }\textbf {\bibinfo {volume} {D77}},\
  \bibinfo {pages} {126007} (\bibinfo {year} {2008})},\ \Eprint
  {http://arxiv.org/abs/0710.3883} {arXiv:0710.3883 [hep-th]} \BibitemShut
  {NoStop}%
\bibitem [{\citenamefont {Kallosh}\ and\ \citenamefont
  {Linde}(2004)}]{Kallosh:2004yh}%
  \BibitemOpen
  \bibfield  {author} {\bibinfo {author} {\bibfnamefont {R.}~\bibnamefont
  {Kallosh}}\ and\ \bibinfo {author} {\bibfnamefont {A.~D.}\ \bibnamefont
  {Linde}},\ }\href {\doibase 10.1088/1126-6708/2004/12/004} {\bibfield
  {journal} {\bibinfo  {journal} {JHEP}\ }\textbf {\bibinfo {volume} {0412}},\
  \bibinfo {pages} {004} (\bibinfo {year} {2004})},\ \Eprint
  {http://arxiv.org/abs/hep-th/0411011} {arXiv:hep-th/0411011 [hep-th]}
  \BibitemShut {NoStop}%
\bibitem [{\citenamefont {Hayashi}\ \emph {et~al.}(2014)\citenamefont
  {Hayashi}, \citenamefont {Matsuda},\ and\ \citenamefont
  {Watari}}]{Hayashi:2014aua}%
  \BibitemOpen
  \bibfield  {author} {\bibinfo {author} {\bibfnamefont {H.}~\bibnamefont
  {Hayashi}}, \bibinfo {author} {\bibfnamefont {R.}~\bibnamefont {Matsuda}}, \
  and\ \bibinfo {author} {\bibfnamefont {T.}~\bibnamefont {Watari}},\
  }\href@noop {} {\  (\bibinfo {year} {2014})},\ \Eprint
  {http://arxiv.org/abs/1410.7522} {arXiv:1410.7522 [hep-th]} \BibitemShut
  {NoStop}%
\bibitem [{\citenamefont {Kachru}\ \emph
  {et~al.}(2003{\natexlab{b}})\citenamefont {Kachru}, \citenamefont {Kallosh},
  \citenamefont {Linde}, \citenamefont {Maldacena}, \citenamefont {McAllister}
  \emph {et~al.}}]{Kachru:2003sx}%
  \BibitemOpen
  \bibfield  {author} {\bibinfo {author} {\bibfnamefont {S.}~\bibnamefont
  {Kachru}}, \bibinfo {author} {\bibfnamefont {R.}~\bibnamefont {Kallosh}},
  \bibinfo {author} {\bibfnamefont {A.~D.}\ \bibnamefont {Linde}}, \bibinfo
  {author} {\bibfnamefont {J.~M.}\ \bibnamefont {Maldacena}}, \bibinfo {author}
  {\bibfnamefont {L.~P.}\ \bibnamefont {McAllister}},  \emph {et~al.},\ }\href
  {\doibase 10.1088/1475-7516/2003/10/013} {\bibfield  {journal} {\bibinfo
  {journal} {JCAP}\ }\textbf {\bibinfo {volume} {0310}},\ \bibinfo {pages}
  {013} (\bibinfo {year} {2003}{\natexlab{b}})},\ \Eprint
  {http://arxiv.org/abs/hep-th/0308055} {arXiv:hep-th/0308055 [hep-th]}
  \BibitemShut {NoStop}%
\bibitem [{\citenamefont {Baumann}\ \emph {et~al.}(2007)\citenamefont
  {Baumann}, \citenamefont {Dymarsky}, \citenamefont {Klebanov}, \citenamefont
  {McAllister},\ and\ \citenamefont {Steinhardt}}]{Baumann:2007np}%
  \BibitemOpen
  \bibfield  {author} {\bibinfo {author} {\bibfnamefont {D.}~\bibnamefont
  {Baumann}}, \bibinfo {author} {\bibfnamefont {A.}~\bibnamefont {Dymarsky}},
  \bibinfo {author} {\bibfnamefont {I.~R.}\ \bibnamefont {Klebanov}}, \bibinfo
  {author} {\bibfnamefont {L.}~\bibnamefont {McAllister}}, \ and\ \bibinfo
  {author} {\bibfnamefont {P.~J.}\ \bibnamefont {Steinhardt}},\ }\href
  {\doibase 10.1103/PhysRevLett.99.141601} {\bibfield  {journal} {\bibinfo
  {journal} {Phys.Rev.Lett.}\ }\textbf {\bibinfo {volume} {99}},\ \bibinfo
  {pages} {141601} (\bibinfo {year} {2007})},\ \Eprint
  {http://arxiv.org/abs/0705.3837} {arXiv:0705.3837 [hep-th]} \BibitemShut
  {NoStop}%
\bibitem [{\citenamefont {Baumann}\ \emph {et~al.}(2008)\citenamefont
  {Baumann}, \citenamefont {Dymarsky}, \citenamefont {Klebanov},\ and\
  \citenamefont {McAllister}}]{Baumann:2007ah}%
  \BibitemOpen
  \bibfield  {author} {\bibinfo {author} {\bibfnamefont {D.}~\bibnamefont
  {Baumann}}, \bibinfo {author} {\bibfnamefont {A.}~\bibnamefont {Dymarsky}},
  \bibinfo {author} {\bibfnamefont {I.~R.}\ \bibnamefont {Klebanov}}, \ and\
  \bibinfo {author} {\bibfnamefont {L.}~\bibnamefont {McAllister}},\ }\href
  {\doibase 10.1088/1475-7516/2008/01/024} {\bibfield  {journal} {\bibinfo
  {journal} {JCAP}\ }\textbf {\bibinfo {volume} {0801}},\ \bibinfo {pages}
  {024} (\bibinfo {year} {2008})},\ \Eprint {http://arxiv.org/abs/0706.0360}
  {arXiv:0706.0360 [hep-th]} \BibitemShut {NoStop}%
\bibitem [{\citenamefont {Chen}\ \emph {et~al.}(2009)\citenamefont {Chen},
  \citenamefont {Hung},\ and\ \citenamefont {Shiu}}]{Chen:2009nk}%
  \BibitemOpen
  \bibfield  {author} {\bibinfo {author} {\bibfnamefont {H.-Y.}\ \bibnamefont
  {Chen}}, \bibinfo {author} {\bibfnamefont {L.-Y.}\ \bibnamefont {Hung}}, \
  and\ \bibinfo {author} {\bibfnamefont {G.}~\bibnamefont {Shiu}},\ }\href
  {\doibase 10.1088/1126-6708/2009/03/083} {\bibfield  {journal} {\bibinfo
  {journal} {JHEP}\ }\textbf {\bibinfo {volume} {0903}},\ \bibinfo {pages}
  {083} (\bibinfo {year} {2009})},\ \Eprint {http://arxiv.org/abs/0901.0267}
  {arXiv:0901.0267 [hep-th]} \BibitemShut {NoStop}%
\bibitem [{\citenamefont {Lebedev}\ \emph {et~al.}(2006)\citenamefont
  {Lebedev}, \citenamefont {Nilles},\ and\ \citenamefont
  {Ratz}}]{Lebedev:2006qq}%
  \BibitemOpen
  \bibfield  {author} {\bibinfo {author} {\bibfnamefont {O.}~\bibnamefont
  {Lebedev}}, \bibinfo {author} {\bibfnamefont {H.~P.}\ \bibnamefont {Nilles}},
  \ and\ \bibinfo {author} {\bibfnamefont {M.}~\bibnamefont {Ratz}},\ }\href
  {\doibase 10.1016/j.physletb.2006.03.046} {\bibfield  {journal} {\bibinfo
  {journal} {Phys.Lett.}\ }\textbf {\bibinfo {volume} {B636}},\ \bibinfo
  {pages} {126} (\bibinfo {year} {2006})},\ \Eprint
  {http://arxiv.org/abs/hep-th/0603047} {arXiv:hep-th/0603047 [hep-th]}
  \BibitemShut {NoStop}%
\bibitem [{\citenamefont {Dudas}\ \emph {et~al.}(2007)\citenamefont {Dudas},
  \citenamefont {Papineau},\ and\ \citenamefont {Pokorski}}]{Dudas:2006gr}%
  \BibitemOpen
  \bibfield  {author} {\bibinfo {author} {\bibfnamefont {E.}~\bibnamefont
  {Dudas}}, \bibinfo {author} {\bibfnamefont {C.}~\bibnamefont {Papineau}}, \
  and\ \bibinfo {author} {\bibfnamefont {S.}~\bibnamefont {Pokorski}},\ }\href
  {\doibase 10.1088/1126-6708/2007/02/028} {\bibfield  {journal} {\bibinfo
  {journal} {JHEP}\ }\textbf {\bibinfo {volume} {0702}},\ \bibinfo {pages}
  {028} (\bibinfo {year} {2007})},\ \Eprint
  {http://arxiv.org/abs/hep-th/0610297} {arXiv:hep-th/0610297 [hep-th]}
  \BibitemShut {NoStop}%
\bibitem [{\citenamefont {Abe}\ \emph {et~al.}(2007)\citenamefont {Abe},
  \citenamefont {Higaki}, \citenamefont {Kobayashi},\ and\ \citenamefont
  {Omura}}]{Abe:2006xp}%
  \BibitemOpen
  \bibfield  {author} {\bibinfo {author} {\bibfnamefont {H.}~\bibnamefont
  {Abe}}, \bibinfo {author} {\bibfnamefont {T.}~\bibnamefont {Higaki}},
  \bibinfo {author} {\bibfnamefont {T.}~\bibnamefont {Kobayashi}}, \ and\
  \bibinfo {author} {\bibfnamefont {Y.}~\bibnamefont {Omura}},\ }\href
  {\doibase 10.1103/PhysRevD.75.025019} {\bibfield  {journal} {\bibinfo
  {journal} {Phys.Rev.}\ }\textbf {\bibinfo {volume} {D75}},\ \bibinfo {pages}
  {025019} (\bibinfo {year} {2007})},\ \Eprint
  {http://arxiv.org/abs/hep-th/0611024} {arXiv:hep-th/0611024 [hep-th]}
  \BibitemShut {NoStop}%
\bibitem [{\citenamefont {Kallosh}\ and\ \citenamefont
  {Linde}(2007)}]{Kallosh:2006dv}%
  \BibitemOpen
  \bibfield  {author} {\bibinfo {author} {\bibfnamefont {R.}~\bibnamefont
  {Kallosh}}\ and\ \bibinfo {author} {\bibfnamefont {A.~D.}\ \bibnamefont
  {Linde}},\ }\href {\doibase 10.1088/1126-6708/2007/02/002} {\bibfield
  {journal} {\bibinfo  {journal} {JHEP}\ }\textbf {\bibinfo {volume} {0702}},\
  \bibinfo {pages} {002} (\bibinfo {year} {2007})},\ \Eprint
  {http://arxiv.org/abs/hep-th/0611183} {arXiv:hep-th/0611183 [hep-th]}
  \BibitemShut {NoStop}%
\bibitem [{\citenamefont {Dudas}\ \emph {et~al.}(2008)\citenamefont {Dudas},
  \citenamefont {Mambrini}, \citenamefont {Pokorski},\ and\ \citenamefont
  {Romagnoni}}]{Dudas:2007nz}%
  \BibitemOpen
  \bibfield  {author} {\bibinfo {author} {\bibfnamefont {E.}~\bibnamefont
  {Dudas}}, \bibinfo {author} {\bibfnamefont {Y.}~\bibnamefont {Mambrini}},
  \bibinfo {author} {\bibfnamefont {S.}~\bibnamefont {Pokorski}}, \ and\
  \bibinfo {author} {\bibfnamefont {A.}~\bibnamefont {Romagnoni}},\ }\href
  {\doibase 10.1088/1126-6708/2008/04/015} {\bibfield  {journal} {\bibinfo
  {journal} {JHEP}\ }\textbf {\bibinfo {volume} {0804}},\ \bibinfo {pages}
  {015} (\bibinfo {year} {2008})},\ \Eprint {http://arxiv.org/abs/0711.4934}
  {arXiv:0711.4934 [hep-th]} \BibitemShut {NoStop}%
\bibitem [{\citenamefont {Dudas}\ \emph {et~al.}(2009)\citenamefont {Dudas},
  \citenamefont {Mambrini}, \citenamefont {Pokorski}, \citenamefont
  {Romagnoni},\ and\ \citenamefont {Trapletti}}]{Dudas:2008qf}%
  \BibitemOpen
  \bibfield  {author} {\bibinfo {author} {\bibfnamefont {E.}~\bibnamefont
  {Dudas}}, \bibinfo {author} {\bibfnamefont {Y.}~\bibnamefont {Mambrini}},
  \bibinfo {author} {\bibfnamefont {S.}~\bibnamefont {Pokorski}}, \bibinfo
  {author} {\bibfnamefont {A.}~\bibnamefont {Romagnoni}}, \ and\ \bibinfo
  {author} {\bibfnamefont {M.}~\bibnamefont {Trapletti}},\ }\href {\doibase
  10.1088/1126-6708/2009/03/011} {\bibfield  {journal} {\bibinfo  {journal}
  {JHEP}\ }\textbf {\bibinfo {volume} {0903}},\ \bibinfo {pages} {011}
  (\bibinfo {year} {2009})},\ \Eprint {http://arxiv.org/abs/0809.5064}
  {arXiv:0809.5064 [hep-th]} \BibitemShut {NoStop}%
\bibitem [{\citenamefont {Li}\ \emph {et~al.}(2014{\natexlab{a}})\citenamefont
  {Li}, \citenamefont {Li},\ and\ \citenamefont {Nanopoulos}}]{Li:2014owa}%
  \BibitemOpen
  \bibfield  {author} {\bibinfo {author} {\bibfnamefont {T.}~\bibnamefont
  {Li}}, \bibinfo {author} {\bibfnamefont {Z.}~\bibnamefont {Li}}, \ and\
  \bibinfo {author} {\bibfnamefont {D.~V.}\ \bibnamefont {Nanopoulos}},\
  }\href@noop {} {\  (\bibinfo {year} {2014}{\natexlab{a}})},\ \Eprint
  {http://arxiv.org/abs/1405.0197} {arXiv:1405.0197 [hep-th]} \BibitemShut
  {NoStop}%
\bibitem [{\citenamefont {Li}\ \emph {et~al.}(2014{\natexlab{b}})\citenamefont
  {Li}, \citenamefont {Li},\ and\ \citenamefont {Nanopoulos}}]{Li:2014xna}%
  \BibitemOpen
  \bibfield  {author} {\bibinfo {author} {\bibfnamefont {T.}~\bibnamefont
  {Li}}, \bibinfo {author} {\bibfnamefont {Z.}~\bibnamefont {Li}}, \ and\
  \bibinfo {author} {\bibfnamefont {D.~V.}\ \bibnamefont {Nanopoulos}},\ }\href
  {\doibase 10.1007/JHEP07(2014)052} {\bibfield  {journal} {\bibinfo  {journal}
  {JHEP}\ }\textbf {\bibinfo {volume} {1407}},\ \bibinfo {pages} {052}
  (\bibinfo {year} {2014}{\natexlab{b}})},\ \Eprint
  {http://arxiv.org/abs/1405.1804} {arXiv:1405.1804 [hep-th]} \BibitemShut
  {NoStop}%
\bibitem [{\citenamefont {Li}\ \emph {et~al.}(2014{\natexlab{c}})\citenamefont
  {Li}, \citenamefont {Li},\ and\ \citenamefont {Nanopoulos}}]{Li:2014lpa}%
  \BibitemOpen
  \bibfield  {author} {\bibinfo {author} {\bibfnamefont {T.}~\bibnamefont
  {Li}}, \bibinfo {author} {\bibfnamefont {Z.}~\bibnamefont {Li}}, \ and\
  \bibinfo {author} {\bibfnamefont {D.~V.}\ \bibnamefont {Nanopoulos}},\ }\href
  {\doibase 10.1007/JHEP11(2014)012} {\bibfield  {journal} {\bibinfo  {journal}
  {JHEP}\ }\textbf {\bibinfo {volume} {1411}},\ \bibinfo {pages} {012}
  (\bibinfo {year} {2014}{\natexlab{c}})},\ \Eprint
  {http://arxiv.org/abs/1407.1819} {arXiv:1407.1819 [hep-th]} \BibitemShut
  {NoStop}%
\bibitem [{\citenamefont {Intriligator}\ and\ \citenamefont
  {Seiberg}(1996)}]{Intriligator:1995au}%
  \BibitemOpen
  \bibfield  {author} {\bibinfo {author} {\bibfnamefont {K.~A.}\ \bibnamefont
  {Intriligator}}\ and\ \bibinfo {author} {\bibfnamefont {N.}~\bibnamefont
  {Seiberg}},\ }\href {\doibase 10.1016/0920-5632(95)00626-5} {\bibfield
  {journal} {\bibinfo  {journal} {Nucl.Phys.Proc.Suppl.}\ }\textbf {\bibinfo
  {volume} {45BC}},\ \bibinfo {pages} {1} (\bibinfo {year} {1996})},\ \Eprint
  {http://arxiv.org/abs/hep-th/9509066} {arXiv:hep-th/9509066 [hep-th]}
  \BibitemShut {NoStop}%
\bibitem [{\citenamefont {Font}\ \emph
  {et~al.}(1990{\natexlab{a}})\citenamefont {Font}, \citenamefont {Ibanez},
  \citenamefont {Lust},\ and\ \citenamefont {Quevedo}}]{Font:1990nt}%
  \BibitemOpen
  \bibfield  {author} {\bibinfo {author} {\bibfnamefont {A.}~\bibnamefont
  {Font}}, \bibinfo {author} {\bibfnamefont {L.~E.}\ \bibnamefont {Ibanez}},
  \bibinfo {author} {\bibfnamefont {D.}~\bibnamefont {Lust}}, \ and\ \bibinfo
  {author} {\bibfnamefont {F.}~\bibnamefont {Quevedo}},\ }\href {\doibase
  10.1016/0370-2693(90)90665-S} {\bibfield  {journal} {\bibinfo  {journal}
  {Phys.Lett.}\ }\textbf {\bibinfo {volume} {B245}},\ \bibinfo {pages} {401}
  (\bibinfo {year} {1990}{\natexlab{a}})}\BibitemShut {NoStop}%
\bibitem [{\citenamefont {Font}\ \emph
  {et~al.}(1990{\natexlab{b}})\citenamefont {Font}, \citenamefont {Ibanez},
  \citenamefont {Lust},\ and\ \citenamefont {Quevedo}}]{Font:1990gx}%
  \BibitemOpen
  \bibfield  {author} {\bibinfo {author} {\bibfnamefont {A.}~\bibnamefont
  {Font}}, \bibinfo {author} {\bibfnamefont {L.~E.}\ \bibnamefont {Ibanez}},
  \bibinfo {author} {\bibfnamefont {D.}~\bibnamefont {Lust}}, \ and\ \bibinfo
  {author} {\bibfnamefont {F.}~\bibnamefont {Quevedo}},\ }\href {\doibase
  10.1016/0370-2693(90)90523-9} {\bibfield  {journal} {\bibinfo  {journal}
  {Phys.Lett.}\ }\textbf {\bibinfo {volume} {B249}},\ \bibinfo {pages} {35}
  (\bibinfo {year} {1990}{\natexlab{b}})}\BibitemShut {NoStop}%
\bibitem [{\citenamefont {Abe}\ \emph {et~al.}(2014{\natexlab{c}})\citenamefont
  {Abe}, \citenamefont {Kobayashi},\ and\ \citenamefont
  {Otsuka}}]{Abe:2014pwa}%
  \BibitemOpen
  \bibfield  {author} {\bibinfo {author} {\bibfnamefont {H.}~\bibnamefont
  {Abe}}, \bibinfo {author} {\bibfnamefont {T.}~\bibnamefont {Kobayashi}}, \
  and\ \bibinfo {author} {\bibfnamefont {H.}~\bibnamefont {Otsuka}},\
  }\href@noop {} {\  (\bibinfo {year} {2014}{\natexlab{c}})},\ \Eprint
  {http://arxiv.org/abs/1409.8436} {arXiv:1409.8436 [hep-th]} \BibitemShut
  {NoStop}%
\bibitem [{\citenamefont {Kim}\ \emph {et~al.}(2005)\citenamefont {Kim},
  \citenamefont {Nilles},\ and\ \citenamefont {Peloso}}]{Kim:2004rp}%
  \BibitemOpen
  \bibfield  {author} {\bibinfo {author} {\bibfnamefont {J.~E.}\ \bibnamefont
  {Kim}}, \bibinfo {author} {\bibfnamefont {H.~P.}\ \bibnamefont {Nilles}}, \
  and\ \bibinfo {author} {\bibfnamefont {M.}~\bibnamefont {Peloso}},\ }\href
  {\doibase 10.1088/1475-7516/2005/01/005} {\bibfield  {journal} {\bibinfo
  {journal} {JCAP}\ }\textbf {\bibinfo {volume} {0501}},\ \bibinfo {pages}
  {005} (\bibinfo {year} {2005})},\ \Eprint
  {http://arxiv.org/abs/hep-ph/0409138} {arXiv:hep-ph/0409138 [hep-ph]}
  \BibitemShut {NoStop}%
\bibitem [{\citenamefont {Fukugita}\ and\ \citenamefont
  {Yanagida}(1986)}]{Fukugita:1986hr}%
  \BibitemOpen
  \bibfield  {author} {\bibinfo {author} {\bibfnamefont {M.}~\bibnamefont
  {Fukugita}}\ and\ \bibinfo {author} {\bibfnamefont {T.}~\bibnamefont
  {Yanagida}},\ }\href {\doibase 10.1016/0370-2693(86)91126-3} {\bibfield
  {journal} {\bibinfo  {journal} {Phys.Lett.}\ }\textbf {\bibinfo {volume}
  {B174}},\ \bibinfo {pages} {45} (\bibinfo {year} {1986})}\BibitemShut
  {NoStop}%
\bibitem [{\citenamefont {Asaka}\ \emph {et~al.}(1999)\citenamefont {Asaka},
  \citenamefont {Hamaguchi}, \citenamefont {Kawasaki},\ and\ \citenamefont
  {Yanagida}}]{Asaka:1999yd}%
  \BibitemOpen
  \bibfield  {author} {\bibinfo {author} {\bibfnamefont {T.}~\bibnamefont
  {Asaka}}, \bibinfo {author} {\bibfnamefont {K.}~\bibnamefont {Hamaguchi}},
  \bibinfo {author} {\bibfnamefont {M.}~\bibnamefont {Kawasaki}}, \ and\
  \bibinfo {author} {\bibfnamefont {T.}~\bibnamefont {Yanagida}},\ }\href
  {\doibase 10.1016/S0370-2693(99)01020-5} {\bibfield  {journal} {\bibinfo
  {journal} {Phys.Lett.}\ }\textbf {\bibinfo {volume} {B464}},\ \bibinfo
  {pages} {12} (\bibinfo {year} {1999})},\ \Eprint
  {http://arxiv.org/abs/hep-ph/9906366} {arXiv:hep-ph/9906366 [hep-ph]}
  \BibitemShut {NoStop}%
\bibitem [{\citenamefont {Asaka}\ \emph {et~al.}(2000)\citenamefont {Asaka},
  \citenamefont {Hamaguchi}, \citenamefont {Kawasaki},\ and\ \citenamefont
  {Yanagida}}]{Asaka:1999jb}%
  \BibitemOpen
  \bibfield  {author} {\bibinfo {author} {\bibfnamefont {T.}~\bibnamefont
  {Asaka}}, \bibinfo {author} {\bibfnamefont {K.}~\bibnamefont {Hamaguchi}},
  \bibinfo {author} {\bibfnamefont {M.}~\bibnamefont {Kawasaki}}, \ and\
  \bibinfo {author} {\bibfnamefont {T.}~\bibnamefont {Yanagida}},\ }\href
  {\doibase 10.1103/PhysRevD.61.083512} {\bibfield  {journal} {\bibinfo
  {journal} {Phys.Rev.}\ }\textbf {\bibinfo {volume} {D61}},\ \bibinfo {pages}
  {083512} (\bibinfo {year} {2000})},\ \Eprint
  {http://arxiv.org/abs/hep-ph/9907559} {arXiv:hep-ph/9907559 [hep-ph]}
  \BibitemShut {NoStop}%
\end{thebibliography}%
\end{document}